\documentclass[ALICE,manyauthors]{cernphprep}
\usepackage[comma,square,numbers,sort&compress]{natbib}

\usepackage[T1]{fontenc} 
\usepackage[utf8]{inputenc}
\usepackage[english]{babel}
\usepackage{lineno}
\usepackage{xspace}
\usepackage{hyperref}
\usepackage{amssymb}
\usepackage{hyperref}
\usepackage{lineno}
\usepackage{booktabs}
\usepackage{multicol}
\usepackage{multirow}
\usepackage{xcolor}
\usepackage{overpic}
\usepackage{tikz}
\usepackage{placeins}
\usepackage{units}
\usepackage{siunitx}[=v2]
\usepackage{color}
\usepackage{float}			% provides the H float modifier option
\usepackage{orcidlink}

\usetikzlibrary{calc}
\usepackage[nodayofweek]{datetime}
\usepackage{enumitem}
\setlist[1]{itemsep=-5pt}

\usepackage[toc]{appendix}
\definecolor{mygray}{gray}{0.8}
\def\centerarc[#1](#2)(#3:#4:#5)% Syntax: [draw options] (center) (initial angle:final angle:radius)
    { \draw[#1] ($(#2)+({#5*cos(#3)},{#5*sin(#3)})$) arc (#3:#4:#5); }

\begin{document}
%%%%%%%%%%%%%%%%%%%%%%%%%%%%%%%%%%%%%%%%%%%%%%%%%%
% These are some new commands that may be useful 
% for paper writing in general. If other newcommands
% are needed for your specific paper, please feel 
% free to add here. 
%
% The currently available commands are organized in: 
% 1) Systems
% 2) Quantities
% 3) Energies and units
% 4) Detectors
% 5) particle species 
%%%%%%%%%%%%%%%%%%%%%%%%%%%%%%%%%%%%%%%%%%%%%%%%%%

% 1) SYSTEMS 
\newcommand{\pp}           {pp\xspace}
\newcommand{\ppbar}        {\mbox{$\mathrm {p\overline{p}}$}\xspace}
\newcommand{\XeXe}         {\mbox{Xe--Xe}\xspace}
\newcommand{\PbPb}         {\mbox{Pb--Pb}\xspace}
\newcommand{\pPb}          {\mbox{p--Pb}\xspace}
\newcommand{\AuAu}         {\mbox{Au--Au}\xspace}
\newcommand{\dAu}          {\mbox{d--Au}\xspace}

% 2) QUANTITIES 
\newcommand{\meanpt}       {$\langle p_{\mathrm{T}}\rangle$\xspace}
\newcommand{\ycms}         {\ensuremath{y_{\rm CMS}}\xspace}
\newcommand{\ylab}         {\ensuremath{y_{\rm lab}}\xspace}
\newcommand{\etarange}[1]  {\mbox{$\left | \eta \right |~<~#1$}}
\newcommand{\yrange}[1]    {\mbox{$\left | y \right |~<~#1$}}
\newcommand {\dNchdeta}     {\ensuremath{\mathrm{d}N_\text{ch}/\mathrm{d}\eta }\xspace}
\newcommand{\dndy}         {\ensuremath{\mathrm{d}N_\mathrm{ch}/\mathrm{d}y}\xspace}
\newcommand{\dndyres}         {\ensuremath{\mathrm{d}N/\mathrm{d}y}\xspace}
\newcommand{\avdndy}        {\ensuremath{\langle\mathrm{d}N_\mathrm{ch}/\mathrm{d}y\rangle}\xspace}
\newcommand{\avdndyres}        {\ensuremath{\langle\mathrm{d}N/\mathrm{d}y\rangle}\xspace}
\newcommand{\dNdy}         {\ensuremath{\mathrm{d}N_\mathrm{ch}/\mathrm{d}y}\xspace}
\newcommand{\Npart}        {\ensuremath{N_\mathrm{part}}\xspace}
\newcommand{\Ncoll}        {\ensuremath{N_\mathrm{coll}}\xspace}
\newcommand{\dEdx}         {\ensuremath{\textrm{d}E/\textrm{d}x}\xspace}
\newcommand{\RpPb}         {\ensuremath{R_{\rm pPb}}\xspace}

% 3) ENERGIES, UNITS
\newcommand{\nineH}        {$\sqrt{\it{s}}~=~0.9$~Te\kern-.1emV\xspace}
\newcommand{\seven}        {$\sqrt{\it{s}}=7$~Te\kern-.1emV\xspace}
\newcommand{\twoH}         {$\sqrt{\it{s}}~=~0.2$~Te\kern-.1emV\xspace}
\newcommand{\twosevensix}  {$\sqrt{\it{s}}~=~2.76$~Te\kern-.1emV\xspace}
\newcommand{\five}         {$\sqrt{\it{s}}~=~5.02$~Te\kern-.1emV\xspace}
\newcommand{\thirteen}         {$\sqrt{\it{s}}=13$~Te\kern-.1emV\xspace}
\newcommand{\twosevensixnn}{$\sqrt{s_{\mathrm{NN}}}~=~2.76$~Te\kern-.1emV\xspace}
\newcommand{\fivenn}       {$\sqrt{s_{\mathrm{NN}}}~=~5.02$~Te\kern-.1emV\xspace}
\newcommand{\LT}           {L{\'e}vy-Tsallis\xspace}
\newcommand{\GeVc}         {Ge\kern-.1emV/$c$\xspace}
\newcommand{\MeVc}         {Me\kern-.1emV/$c$\xspace}
\newcommand{\GeVmass}      {GeV/$c^2$\xspace}
\newcommand{\MeVmass}      {MeV/$c^2$\xspace}
\newcommand{\lumi}         {\ensuremath{\mathcal{L}}\xspace}

% 4) DETECTORS 
\newcommand{\ITS}          {\rm{ITS}\xspace}
\newcommand{\TOF}          {\rm{TOF}\xspace}
\newcommand{\ZDC}          {\rm{ZDC}\xspace}
\newcommand{\ZDCs}         {\rm{ZDCs}\xspace}
\newcommand{\ZNA}          {\rm{ZNA}\xspace}
\newcommand{\ZNC}          {\rm{ZNC}\xspace}
\newcommand{\SPD}          {\rm{SPD}\xspace}
\newcommand{\SDD}          {\rm{SDD}\xspace}
\newcommand{\SSD}          {\rm{SSD}\xspace}
\newcommand{\TPC}          {\rm{TPC}\xspace}
\newcommand{\TRD}          {\rm{TRD}\xspace}
\newcommand{\VZERO}        {\rm{V0}\xspace}
\newcommand{\VZEROA}       {\rm{V0A}\xspace}
\newcommand{\VZEROC}       {\rm{V0C}\xspace}
\newcommand{\Vdecay} 	   {\ensuremath{V^{0}}\xspace}

% 4) PARTICLE SPECIES 
\newcommand{\ee}           {\ensuremath{e^{+}e^{-}}} 
\newcommand{\pip}          {\ensuremath{\pi^{+}}\xspace}
\newcommand{\pim}          {\ensuremath{\pi^{-}}\xspace}
\newcommand{\kap}          {\ensuremath{\rm{K}^{+}}\xspace}
\newcommand{\kam}          {\ensuremath{\rm{K}^{-}}\xspace}
\newcommand{\pbar}         {\ensuremath{\rm\overline{p}}\xspace}
\newcommand{\kzero}        {\ensuremath{{\rm K}^{0}_{\rm{S}}}\xspace}
\newcommand{\lmb}          {\ensuremath{\Lambda}\xspace}
\newcommand{\almb}         {\ensuremath{\overline{\Lambda}}\xspace}
\newcommand{\Om}           {\ensuremath{\Omega^-}\xspace}
\newcommand{\Mo}           {\ensuremath{\overline{\Omega}^+}\xspace}
\newcommand{\X}            {\ensuremath{\Xi^-}\xspace}
\newcommand{\Ix}           {\ensuremath{\overline{\Xi}^+}\xspace}
\newcommand{\Xis}          {\ensuremath{\Xi^{\pm}}\xspace}
\newcommand{\Oms}          {\ensuremath{\Omega^{\pm}}\xspace}
\newcommand{\xizero}{\ensuremath{\Xi(1530)^{0}}\xspace}
\newcommand{\barxizero}{\overline{\Xi}(1530)^{0}}
\newcommand{\sigmapm}{\ensuremath{\Sigma(1385)^{\pm}}\xspace}
\newcommand{\sigmap}{\ensuremath{\Sigma(1385)^{+}}\xspace}
\newcommand{\sigmam}{\ensuremath{\Sigma(1385)^{-}}\xspace}
\newcommand{\barsigmapm}{\ensuremath{\overline{\Sigma}(1385)^{\pm}}\xspace}
\newcommand{\barsigmap}{\ensuremath{\overline{\Sigma}(1385)^{+}}\xspace}
\newcommand{\barsigmam}{\ensuremath{\overline{\Sigma}(1385)^{-}}\xspace}
\newcommand{\cms}{\mathrm{\sqrt{\it{s}}}}
\newcommand{\snn}{\sqrt{s_\mathrm{NN}}}
\newcommand{\pt}{p_\mathrm{T}}
\newcommand{\dndeta}{\mathrm{d}N_\mathrm{ch}/\mathrm{d}\eta}
\newcommand{\avdndeta}{\ensuremath{\langle\dndeta\rangle}\xspace}
\newcommand{\kstar}{\mathrm{K^{*}(892)^0}}
\newcommand{\phimeson}{\mathrm{\phi(1020)}}
\newcommand{\mband}{\mathrm{MB}_\mathrm{AND}}
\newcommand{\mbandg}{\mathrm{MB}_\mathrm{AND>0}}
\newcommand{\dcar}{\mathrm{DCA}_r}
\newcommand{\inelg}{\mathrm{INEL>0}}
\newcommand{\pythia}{\mathrm{PYTHIA~8~Monash~2013}}
\newcommand{\levi}{L\'evy--Tsallis\xspace}
\newcommand{\pythiaDef}{\mathrm{PYTHIA\,\,8}\xspace}

\newcommand{\dd}     {\ensuremath{\mathrm{d}}\xspace}
\newcommand{\mt}     {\mbox{\ensuremath{m_{\mathrm{T}}}\xspace}}

%%%%%%%%%%%%%%%  Title page %%%%%%%%%%%%%%%%%%%%%%%%
\begin{titlepage}
% the dates below correspond to CERN approval
% please don't touch: EB chairs will take care
\PHyear{2023}       % required, will be obtained from CERN
\PHnumber{172}      % required, will be obtained from CERN
\PHdate{21 August}  % required, will be obtained from CERN
%%%%%%%%%%%%%%%%%%%%%%%%%%%%%%%%%%%%%%%%%%%%%%%%%%%%

%%% Put your own title + short title here:
\title{Multiplicity-dependent production of $\pmb{\sigmapm}$ and $\pmb{\xizero}$ \\ in pp collisions at $\pmb{\cms}$~=~\unit[13]{TeV}}
\ShortTitle{$\sigmapm$ and $\xizero$ production in pp collisions at $\cms$~=~\unit[13]{TeV}}   % appears on left page headers

%%% Do not change the next lines
\Collaboration{ALICE Collaboration\thanks{See Appendix~\ref{app:collab} for the list of collaboration members}}
\ShortAuthor{ALICE Collaboration} % appears on right page headers, do not change

\begin{abstract}
The production yields of the $\sigmapm$ and $\xizero$ resonances are measured in pp collisions at \thirteen with ALICE. The measurements are performed as a function of the charged-particle multiplicity $\avdndeta$, which is related to the energy density produced in the collision. The results include transverse momentum ($\pt$) distributions, $\pt$-integrated yields, mean transverse momenta of $\sigmapm$ and $\xizero$, as well as ratios of the $\pt$-integrated resonance yields relative to yields of other hadron species. The $\sigmapm/\pi^{\pm}$ and $\xizero/\pi^{\pm}$ yield ratios are consistent with the trend of the enhancement of strangeness production from low to high multiplicity pp collisions, which was previously observed for strange and multi-strange baryons. 
The yield ratio between the measured resonances and the long-lived baryons with the same strangeness content exhibits a hint of a mild increasing trend at low multiplicity, despite too large uncertainties to exclude the flat behaviour.
The results are compared with predictions from models such as EPOS-LHC and $\pythiaDef$ with Rope shoving. The latter provides the best description of the multiplicity dependence of the $\sigmapm$ and $\xizero$ production in pp collisions at $\cms$~=~\unit[13]{TeV}.

\end{abstract}
\end{titlepage}

\setcounter{page}{2} %please do not remove this line

%%%%%%%%%%%%%%%%%%%%%%%%%%%%%%%%
% begin main text
%%%%%%%%%%%%%%%%%%%%%%%%%%%%%%%%

\section{Introduction}

Quantum chromodynamics (QCD)~\cite{Collins:1974ky,Cabibbo:1975ig} predicts an extreme state of nuclear matter at high temperature and energy density where quarks and gluons are not confined into hadrons, the quark--gluon plasma (QGP). These conditions are achieved in ultrarelativistic heavy-ion collisions~\cite{STAR:2005gfr,PHENIX:2004vcz,BRAHMS:2004adc,PHOBOS:2004zne,Busza:2018rrf,ALICE:2022wpn} like those at the Relativistic Heavy Ion Collider (RHIC) and the Large Hadron Collider (LHC). As the system created during the collision evolves, the QGP matter cools down and a transition to a hadron gas occurs when the pseudo-critical temperature is reached. During the subsequent hadronic phase, inelastic scatterings among hadrons a priori stop at the chemical freeze-out and elastic interactions cease at the later time of kinetic freeze-out. 

The measurement of the production of strange hadrons has an important role in the study of the QGP properties ~\cite{Rafelski:1982pu,Rafelski:1980rk,Koch:1986ud}. A large abundance of strange hadrons in nucleus--nucleus (\mbox{AA}\xspace) collisions relative to proton--proton (\pp) collisions has manifested itself, without significant initial--energy or volume dependence from RHIC to LHC energies~\cite{Andersen:1998vu,Abelev:2007xp,ABELEV:2013zaa}. Strangeness production in central heavy-ion collisions is described in the frame of statistical hadronisation models utilising a grand canonical formulation and assuming a hadron gas in thermal and chemical equilibrium at the chemical freeze-out stage~\cite{Cleymans:1992zc,Braun-Munzinger:2003pwq,Cleymans:2006xj,Torrieri:2004zz,Andronic:2017pug}. Meanwhile, a continuous enhancement of strange particles relative to pions has been observed with increasing number of charged particles produced in the final state from \pp, \pPb to peripheral \PbPb collisions~\cite{ALICE:2017jyt}. The statistical model with strangeness canonical suppression~\cite{Redlich:2001kb,Cleymans:2020fsc} and the core-corona superposition model~\cite{Becattini:2008yn,Aichelin:2008mi} predict a multiplicity dependence of strangeness production in small systems. 

Hadron resonances are powerful tools to study the properties of the late hadronic phase in ultrarelativistic heavy-ion collisions since the duration of such a phase is of the same order of magnitude as the resonance lifetimes (few fm/\textit{c})~\cite{Markert:2008jc}. Such resonances are influenced by interactions in the hadronic phase. If resonances decay before the kinetic freeze-out, the resonance yield reconstructed from the kinematics of the decay particles should decrease relative to the primordial resonance yield due to rescattering of the decay particles. Conversely, pseudo-elastic scattering of long-lived hadrons occurring after chemical freeze-out can generate resonances and potentially increase the observed yield. The balance between rescatterings and regeneration depends on the scattering cross sections of the decay products, the density of the produced hadron gas, the lifetime of the resonances, and the hadronic phase duration~\cite{Markert:2008jc}.
Such a hot and dense medium usually cannot be expected to be produced in \pp collisions, in contrast to ultra-relativistic heavy-ion collisions. 
However, recent measurements in high-multiplicity \pp collisions ~\cite{ATLAS:2015hzw,CMS:2016fnw,ALICE:2021nir} showed some features resembling those observed in heavy-ion collisions, which can be understood as due to the collective expansion of the medium.
Thus, the study of multiplicity-dependent resonance production in \pp collisions may provide insight into the role of the hadronic interactions in small systems~\cite{Sjostrand:2020gyg,Bierlich:2021poz,Acharya:2019bli}.

The $\sigmapm$ and $\xizero$ baryons are good candidates for the study of single and double-strange resonances with different lifetimes. These baryonic resonances were previously measured in inelastic pp collisions at $\cms$~=~\unit[7]{TeV}~\cite{ALICE:2014zxz} and as a function of the charged-particle multiplicity in p--Pb collisions at $\snn$~=~\unit[5.02]{TeV}~\cite{ALICE:2017pgw}. 
The present measurements in pp collisions at \thirteen extend our knowledge on the production of \sigmapm and \xizero at different centre-of-mass energy and provide insight, for the first time, into the production of these baryonic resonances as a function of the charged-particle multiplicity in pp collisions.

\section{Experimental setup}

A detailed description of the ALICE detector can be found in~\cite{ALICE:2014sbx,ALICE:2022wpn}, where the configuration in place during the Run 1 period of the LHC (2009--2013) is discussed. This configuration is essentially valid also for the Run 2 (2015--2018) when the data used in this analysis were collected. The main detectors used for the measurement of $\sigmapm$ (also denoted as $\Sigma^{*\pm}$ in the following) and $\xizero$ (denoted as $\Xi^{*0}$) reported here are briefly discussed below.

The V0 detector~\cite{Cortese:2004aa,ALICE:2013axi} consists of two arrays (V0A and V0C) of 32 scintillating counters each. The V0A (V0C) is located at a distance of \unit[329]{cm} (\unit[$-88$]{cm}) away from the nominal interaction point ($z=0$) along the beam line that defines the $z$-axis in the coordinate system. The V0A (V0C) covers the pseudorapidity range $2.8<\eta<5.1$ ($-3.7<\eta<-1.7$) and the full azimuth. 
It is used for triggering, for rejection of beam-induced background events, and for the determination of the multiplicity classes by measuring the sum of the signals from V0A and V0C forming the V0M signal.

The Inner Tracking System (ITS)~\cite{ALICE:2014sbx} is composed of six silicon layers and is the innermost detector of ALICE. The ITS is used for charged track reconstruction, and in particular to provide high-precision points in the vicinity of the primary vertex of the collision. The first two layers of the ITS consist of the Silicon Pixel Detector (SPD), located at an average radial distance $\it{r}$ of 3.9 cm and 7.6 cm from the beam line and covering the pseudorapidity ranges $|\eta|<1.4$ and $|\eta|<2.0$ and the full azimuth around the interaction point. The SPD is used to reconstruct short track segments that are called tracklets and to determine the primary vertex of the collision. Beyond the SPD, there are two layers of Silicon Drift Detectors (SDD) and two layers of Silicon Strip Detectors (SSD), with the outermost layer having a radius $r=43$~cm.

The Time Projection Chamber (TPC)~\cite{ALICE:2014sbx}, located just outside the ITS, is a \unit[90]{$\mathrm{m^3}$} cylindrical drift chamber filled with a gas mixture and has a large number of readout channels (557568~\cite{Alme:2010ke}). The radial and the longitudinal dimensions of the TPC are about 85 < $r$ < 250 cm and -250 < $z$ < 250 cm, respectively. The TPC covers the pseudorapidity range $|\eta|<0.9$ and the full azimuthal angle. It provides excellent momentum and spatial resolution for the reconstruction of charged-particle tracks. Besides its tracking capability, the TPC is used for particle identification by measuring the specific ionisation energy loss (d$E$/d$x$) in the gas.

\section{Data analysis}
\label{sec:dataanalysis}

The data sample used in this study was collected with the ALICE detector during the LHC Run~2 (2015--2018) in pp collisions at $\cms$~=~\unit[13]{TeV} with a minimum bias (MB) trigger, which selects inelastic collisions based on the requirement of coincident signals in the V0A and V0C detectors. In addition to the MB trigger requirement, at least one primary charged-particle track in the $|\eta|<1$ range is required in the offline event selection ($\inelg$), to minimise the fraction of diffractive events in the sample~\cite{ALICE:2020swj}. 
The transverse momentum ($\pt$) thresholds of the V0 and SPD detectors are around 50 MeV/$\it{c}$~\cite{ALICE:2015olq,Cheynis:2006nd}. 
Events with pileup of collisions occurring in different bunch crossings (out-of-bunch pileup) within the V0 readout time are rejected based on the timing information of the V0. Events with collision pileup in the same bunch crossing (in-bunch pileup) are removed based on the presence multiple primary vertices reconstructed from SPD tracklets~\cite{ALICE:2015olq}.
Further background events are rejected by using the correlation between the number of hits and the number of tracklets in the SPD. 
Events having a primary vertex (PV) reconstructed from global tracks (at least ITS+TPC information), within the range of $\pm$\unit[10]{$\mathrm{cm}$} with respect to the nominal interaction point along the beam axis are considered. 
The total number of minimum bias triggered pp events analysed is $1.82\times10^9$ corresponding to an integrated luminosity of $\mathcal{L}_\mathrm{int}$~=~\unit[31.5] {$\mathrm{nb}^{-1}$}~\cite{ALICE-PUBLIC-2021-005}.

The $\inelg$ data sample is split into several event classes denoted with Roman numerals by measuring the event activity via the total charge deposited in both V0 detectors, see details in Ref~\cite{ALICE:2019avo}. Table~\ref{tab:multtable} presents the event classes used in this analysis, their corresponding percentile of the $\inelg$ class, and their mean charged-particle multiplicity density $\avdndeta$ measured in $|\eta|<0.5$. Note that some event classes (e.g. I, II, and III) are merged in this analysis (e.g. I+II+III) to increase the statistical significance. The detailed information on $\avdndeta$ distributions and values in each event class is reported in~\cite{ALICE:2020swj}. 

In addition to the study of the multiplicity dependence of particle production in the $\inelg$ data sample~\cite{ALICE:2020swj}, a separate analysis based on the data from inelastic events (INEL)~\cite{ALICE:2015olq} is carried out. This inelastic (INEL) analysis differs from the $\inelg$ only by the event selection based on the MB trigger, that does not request the condition that at least one primary charged-particle track in the $|\eta|<1$ range is present.
It implies that the corresponding event normalisation and its corrections differ; for INEL, the correction on the event normalisation is obtained from the ratio of the ALICE visible cross section to the \textit{total} inelastic cross section~\cite{ALICE-PUBLIC-2021-005,ALICE:2020jsh}

\begin{table}[!tb]
	\centering
	\caption{The event classes~\cite{ALICE:2019avo} used in this analysis, their corresponding multiplicity percentile P $\left(\inelg\right)(\%)$ and the mean charged-particle multiplicity, $\avdndeta$~\cite{ALICE:2019avo,ALICE:2020jsh}. The $\avdndeta$ in inelastic events is $5.31\pm0.18$~\cite{ALICE:2015qqj}.} 
	\label{tab:multtable}
	\begin{tabular}{ c c c }

	\hline
	Event Class & P $\left(\inelg\right)(\%)$ & $\avdndeta$ \\ 
	\hline

	I+II+III & ~~~~~~~0--9.15                        &  $18.67\pm0.20$ \\ 
	IV+V+VI & ~9.15--27.50              & $11.46\pm0.13$ \\ 
	VII+VIII & ~27.50--46.12             & ~~$7.13\pm0.08$ \\ 
	IX & ~46.12--65.53            & ~~$4.49\pm0.05$ \\ 
    X & ~65.53--100.00            & ~~$2.54\pm0.03$ \\ 
\hline

	\end{tabular}
\end{table}

\subsection{Reconstruction of $\sigmapm$ and $\xizero$}

The properties of the particles involved in this analysis and the decay modes used for their reconstruction together  with the branching ratios are reported in Table~\ref{tab:resonances}.

\begin{table}[!b]
\centering
\caption{
Quark content, mass, width, lifetime, decay channel used in this analysis and corresponding branching ratio for the \sigmapm and \xizero resonances as well as for the $\Xi^{-}$ and $\Lambda$~\cite{ParticleDataGroup:2022pth} hyperons. Antiparticles are not listed for conciseness.}
\label{tab:resonances}
\begin{tabular}{lllllll}
\hline
\multirow{2}{*}{Baryon} & Valence & Mass & Width & $\it{c}\rm{\tau}$ & \multirow{2}{*}{Decay channel} & B.R. \\ 
 & quarks  & (\MeVmass) & (\MeVmass) & (fm) & & (\%) \\
\hline
\sigmap & uus & $1382.8 \pm 0.4 $ & $36.0 \pm 0.7$ & ~~$5.5\pm0.1$ &  ~~$\Lambda + \pi^{+}$ & $87.0 \pm 1.5$ \\ 
\sigmam & dds & $1387.2 \pm 0.5 $ & $39.4 \pm 2.1$ & ~~$5.0\pm0.3$ & ~~$\Lambda + \pi^{-}$ & $87.0 \pm 1.5$ \\ \hline
\xizero & uss & $1531.8 \pm 0.3$ & ~~$9.1 \pm 0.5$ & $22.0\pm1.0$ & $\Xi^{-} + \pi^{+}$ & $66.7$ \\ \hline
$\Xi^{-}$ & dss & $1321.7 \pm 0.1$ & & $4.91\times10^{13}$  & ~~$\Lambda + \pi^{-}$ & $99.9$ \\ \hline
$\Lambda$ & uds & $1115.7$ & & $7.89\times10^{13}$ & ~~~$\rm{p} + \pi^{-}$ & $63.9 \pm 0.5$ \\ \hline
\end{tabular}
\end{table}

The charged $\sigmapm$ and the $\xizero$ are reconstructed via their hadronic decay channels. The two charged states $\Sigma(1385)^{+}$, $\Sigma(1385)^{-}$ and their anti-particles ($\overline{\Sigma}(1385)^{-}$, and $\overline{\Sigma}(1385)^{+}$) are separately reconstructed via
\begin{equation}
\begin{aligned}
\sigmap & \rightarrow \Lambda + \pi^{+} \rightarrow (\mathrm{p} + \pi^{-}) + \pi^{+} \\
\barsigmam & \rightarrow \overline{\Lambda} + \pi^{-} \rightarrow (\overline{\mathrm{p}} + \pi^{+}) + \pi^{-} \\
\sigmam & \rightarrow \Lambda + \pi^{-} \rightarrow (\mathrm{p} + \pi^{-}) + \pi^{-} \\
\barsigmap & \rightarrow \overline{\Lambda} + \pi^{+} \rightarrow (\overline{\mathrm{p}} + \pi^{+}) + \pi^{+}\quad.
\end{aligned}
\end{equation}
Likewise, $\xizero$ and its antiparticle ($\barxizero$) are reconstructed via
\begin{equation}
\begin{aligned}
\xizero & \rightarrow \Xi^{-} + \pi^{+} \rightarrow (\Lambda  + \pi^{-}) + \pi^{+} \rightarrow [(\mathrm{p} + \pi^{-}) + \pi^{-}] + \pi^{+} \\
\barxizero & \rightarrow \overline{\Xi}^{+} + \pi^{-} \rightarrow (\overline{\Lambda}  + \pi^{+}) + \pi^{-} \rightarrow [(\overline{\mathrm{p}} + \pi^{+}) + \pi^{+}] + \pi^{-} \quad.
\end{aligned}
\end{equation}

\begin{figure}[!h]
\begin{subfigure}{0.495\textwidth}
    \includegraphics[width=\linewidth]{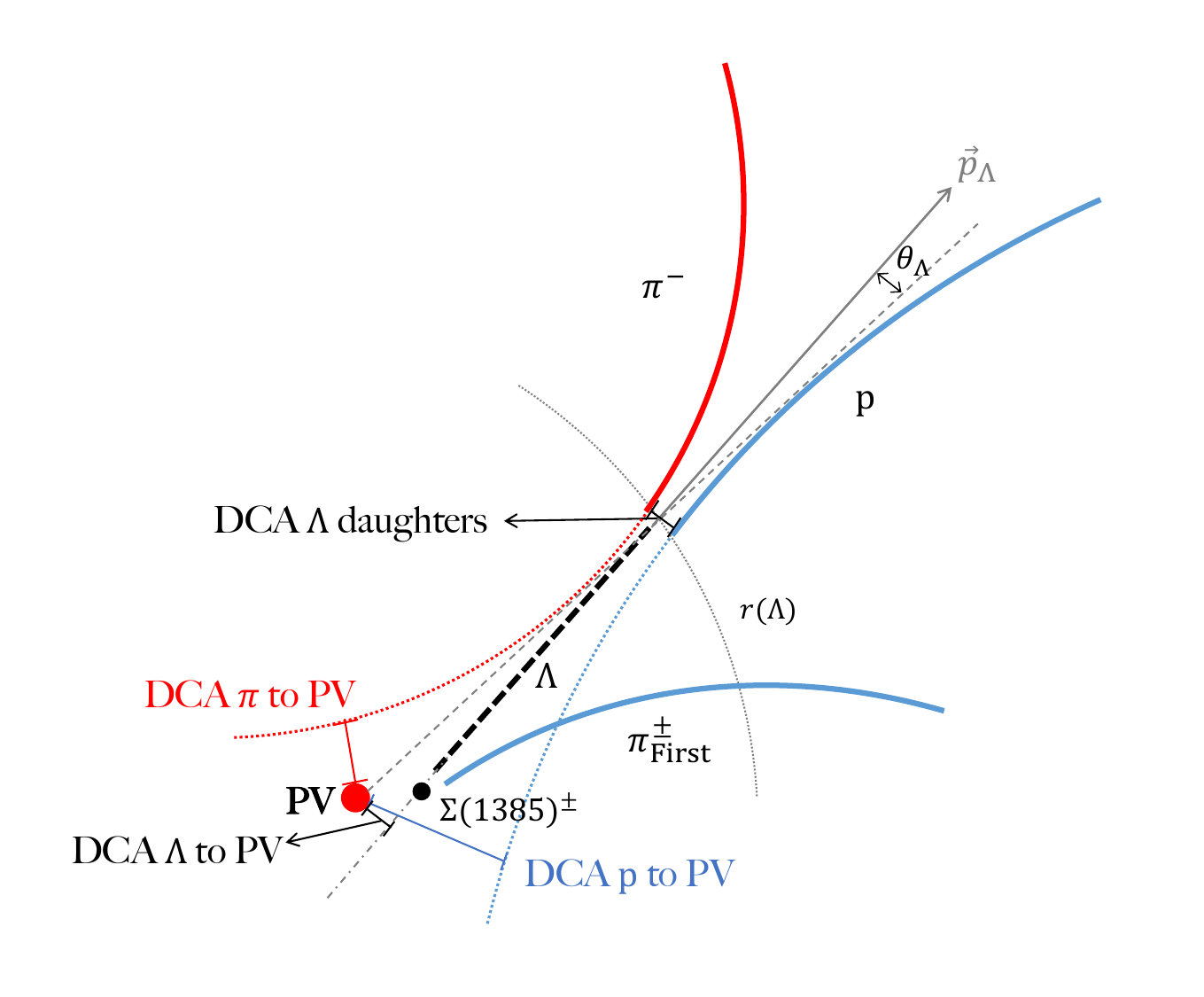}
  \caption{$\sigmapm$}
\end{subfigure}
\begin{subfigure}{0.495\textwidth}
    \includegraphics[width=\linewidth]{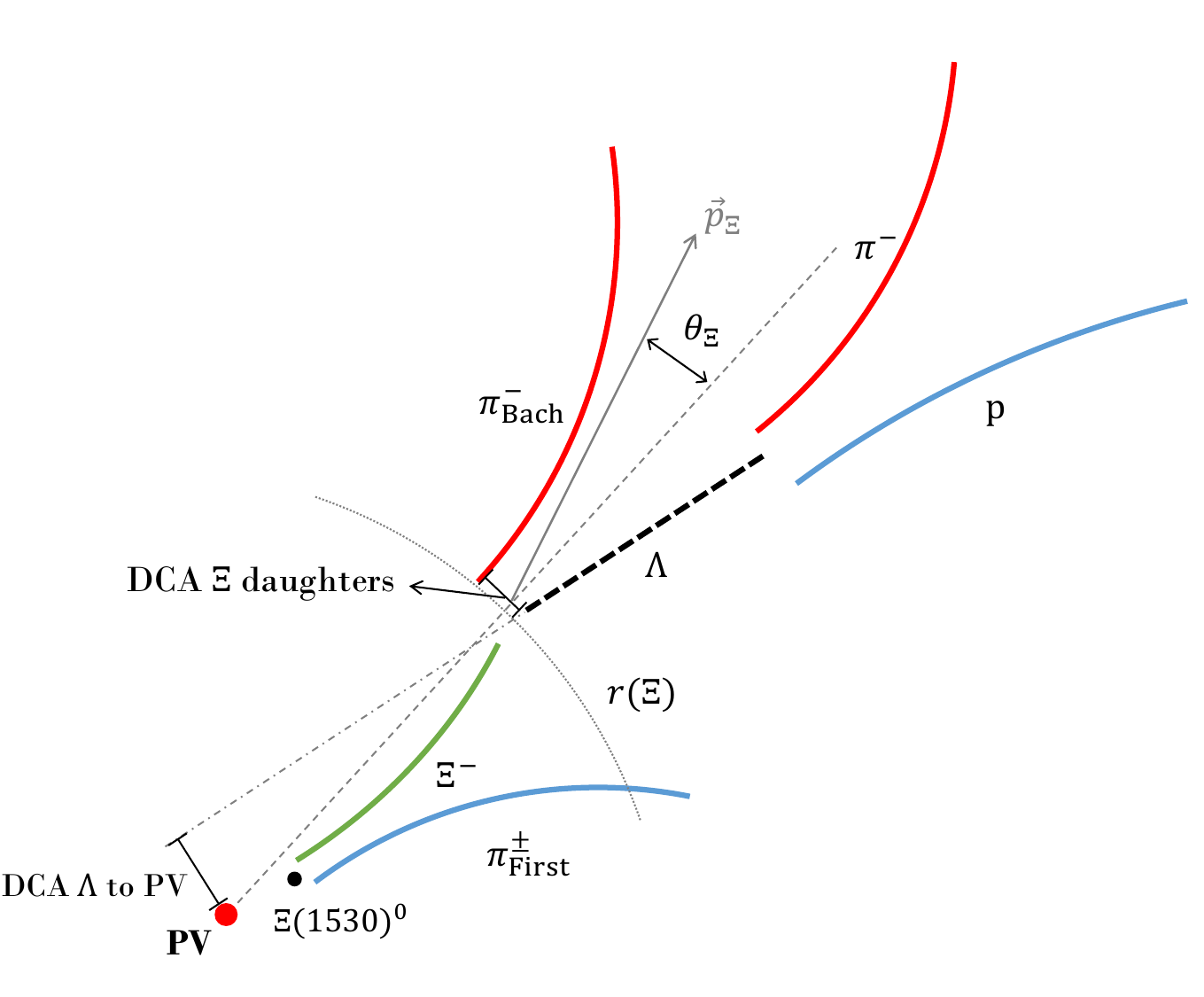}
  \caption{$\xizero$}
\end{subfigure}
\caption{Sketch of the decay modes of $\sigmapm$ and $\xizero$ and depiction of the relevant variables employed for the selection of displaced decay topologies. The distance between the decay vertex (black circle) of the resonances and the PV (red circle) is inflated for clarity, that is, to separate such vertices normally just a few dozen femtometers away from one another.}
\label{fig:decaymode}
\end{figure}

The decays are schematically presented in Fig.~\ref{fig:decaymode} which illustrates the major variables used in the selection of decay topologies on a magnified scale for clarity.

The $\pt$-differential yields in inelastic \pp collisions are calculated using the following equation:

\begin{equation}
  \frac{1}{N_\mathrm{event}} \frac{\mathrm{d}^2N}{\mathrm{d}p_\mathrm{T}\mathrm{d}y} = \frac{\epsilon_{\rm{Trig}} \times \epsilon_{\rm{Vertex}}}{N_{\rm{event,raw}} }\frac{N_{\rm{raw}} \times f_{\rm{S.L.}}}{\Delta\pt \Delta y }\frac{1}{A\times\epsilon_{\rm{rec}}\times\rm{B.R.}}
  \label{eq:analysis}
\end{equation}

where $N_\mathrm{event,raw}$ is the total number of analysed events after the online trigger and the offline selections, $N_{\rm{raw}}$ is the raw yield of the particles extracted in each $\pt$ and rapidity interval, with widths of $\Delta\pt$ and $\Delta y$. 
The factor $\it{A}$ is the acceptance of the detector, ${\epsilon}_{\rm{rec}}$ is the resonance reconstruction efficiency, and $\rm{B.R.}$ is the branching ratio of the decay used for the \sigmapm and \xizero reconstruction. $\epsilon_{\rm{Trig}}$ is the trigger efficiency, $\epsilon_{\rm{Vertex}}$ is the vertex selection efficiency, and both correct the number of events. $f_{\rm{S.L.}}$ is the $p_{\rm{T}}$-dependent signal loss correction factor correcting the raw signal yields.
The extraction of the raw yield, $N_{\rm{raw}}$, is discussed in Section~\ref{subsec:sigext}. 
The correction factors, such as the $\pt$-dependent $A \times {\epsilon}_{\rm{rec}} \times \rm{B.R.}$, the multiplicity-dependent $\epsilon_{\rm{Trig}}$ and $\epsilon_{\rm{Vertex}}$, and $f_{\rm{S.L.}}$ which is dependent on both $\pt$ and multiplicity, are discussed in Section~\ref{subsec:corrections}.

\subsection{Track and topological selections}

Due to the very short lifetime of the strong decaying \sigmapm and \xizero baryons, pions and hyperons originating from the primary vertex are considered for the reconstruction of the resonances.
Pions from the primary vertex are required to have $\pt>0.15$ GeV and to be located in the pseudorapidity range $|\eta|<0.8$ to avoid edge effects in the TPC acceptance~\cite{ALICE:2014sbx}. To ensure a good track reconstruction quality, primary tracks are required to cross at least 70 out of 159 TPC readout rows with a normalised $\chi^2$ ($\chi^2$ per TPC space point) lower than 4. In addition, tracks are required to have a ratio of crossed readout rows, $N_{\rm{crossed}}$, to the number of findable clusters, $N_{\rm{findablecluster}}$, in the TPC larger than 80\%. Primary tracks are also required to have at least one hit in the SPD. The $\chi^2_{\rm{ITS - TPC}}$, calculated by comparing with combined ITS+TPC track parameters to those obtained only from the TPC and constrained to the interaction point, is required to be lower than 36. $\Xi^{-}$ and $\Lambda$ baryons produced in the resonance decays are long lived and they are reconstructed from their decay particles produced in secondary vertices displaced from the primary vertex. The decay particles produced in these secondary vertices are selected among tracks with $|\eta|<0.8$ based on a looser selection with respect to the one for the primary tracks. They are required to cross at least 60 TPC readout rows, while no request on ITS hits is applied. Finally, the selected pion and proton candidates are identified by requiring that the specific ionisation energy loss, $\rm{d}\it{E}/\rm{d}\it{x}$, measured in the TPC lies within three standard deviations ($\sigma_{\rm{TPC}}$) from the specific energy loss expected for pions or protons.

The first emitted pion ($\pi^{\pm}_{\,\rm{First}}$ in Fig.\ref{fig:decaymode}) tracks appear as if they originate from the primary vertex (PV), so they are selected using the condition that the distance of closest approach ($\mathrm{DCA}$) to the primary vertex along the beam axis ($\mathrm{DCA}_z$) should be lower than 2 cm and the DCA in the transverse plane ($\mathrm{DCA}_r$) lower than $0.0105+0.035 \pt^{-1.1}$ cm, with $\pt$ in units of \GeVc. The primary track selection criteria, which are the standard criteria used in ALICE analyses, are summarised in Table~\ref{tab:trackcut}, along with those for secondary tracks.

\begin{table}[!bt]
	\centering
	\caption{Summary of the track selection criteria applied to primary and secondary tracks. The unit of $\pt$ in the $\mathrm{DCA}_r$ to PV formula is \GeVc.} 
	\label{tab:trackcut}
\begin{tabular}{ l | l }
\hline
\multicolumn{2}{l}{Secondary track selection} \\
\hline
$|\eta|$ & $<0.8$ \\
$N_{\rm{crossed}}$ & $>60$ \\
TPC $\rm{d}\it{E}/\rm{d}\it{x}$ ($\sigma$) & $<3$ \\ 
\hline
\multicolumn{2}{l}{Primary track selection} \\
\hline
$|\eta|$ & $<0.8$ \\
$\pt$ (\GeVc) & $>0.15$ \\
$N_{\rm{crossed}}$ & $>70$ \\
$N_{\rm{crossed}}/N_{\rm{findablecluster}}$ & $>0.8$ \\
$\chi^2$/cluster in TPC & $<4$ \\
$\chi^2$ of ITS-TPC track fit & $<36$ \\
$\mathrm{DCA}_z$ to PV (cm) & $<2$ \\
$\mathrm{DCA}_r$ to PV (cm) & $<0.0105+0.035 \pt^{-1.1}$ \\
number of SPD points & $\geq 1$ \\
TPC $\rm{d}\it{E}/\rm{d}\it{x}$ ($\sigma$) & $<3$ \\ 
\hline
\end{tabular}
\end{table}

The secondary vertices of $\Lambda$ and $\Xi^{-}$ are reconstructed via their decay mode into p+$\pi^{-}$ and $\Lambda$+$\pi^{-}$ (and charge conjugates) respectively, by applying a similar strategy as the one used in~\cite{ALICE:2020jsh,ALICE:2019avo,ALICE:2014zxz,ALICE:2017pgw}. The applied geometrical selections on the displaced decay-vertex topology are summarised in Table~\ref{tab:topological cut}.

The $\lmb$($\almb$) from $\sigmapm$ decays are selected if the DCA between the two daughter tracks (DCA $\Lambda$ daughters in Fig.\ref{fig:decaymode}(a)) is smaller than 1.6 cm and the DCA of the $\lmb$ in the $xy$-plane to the PV ($r(\Lambda)$ in Fig.\ref{fig:decaymode}(a)) is larger than 0.02 cm to ensure that those tracks are not primary charged particles coming from the PV. In addition, the DCA of $\Lambda$ to the PV is required to be lower than 0.3 cm to reject the $\Lambda$ baryons from $\Xi^{-}$ or $\overline{\Xi}^{+}$~\cite{ALICE:2014zxz}. The invariant mass of the $m_\mathrm{p\pi^{-}}$  pair is required to be within $\pm10$ MeV/$c^2$ with respect to the world-average $\Lambda$ mass value~\cite{ParticleDataGroup:2022pth}, i.e. within a mass window which is at least about 4 times the reconstructed mass resolution for $\Lambda$~\cite{ALICE:2013wgn}. 
The cosine of the pointing angle ($\theta_{\Lambda}$) between the direction of the momentum of the $\Lambda$ and the line connecting the secondary to the primary vertex is required to be larger than 0.98 to reject any potential secondary $\Lambda$ from other particle decays.

\begin{table}[!bt]
	\centering
	\caption{Summary of the selection criteria for $\sigmapm$ and $\xizero$ candidates. See the text and Fig.~\ref{fig:decaymode} for the details.} 
	\label{tab:topological cut}
\begin{tabular}{l|l|l}
\hline
Selection criteria & $\sigmapm$ & $\xizero$ \\
\hline
DCA $\Lambda$ daughters (cm) &  $<1.6$ & $<1.4$ \\
DCA $\Lambda$ to PV (cm) & $<0.3$ & $>0.07$ \\
DCA $\pi$ to PV (cm) & $>0.05$ & $>0.05$ \\
DCA $\rm{p}$ to PV (cm) & $>0.05$ & $>0.05$ \\
$\cos{\theta_\Lambda}$ & $>0.98$ & $>0.97$ \\ 
$r$($\Lambda$) (cm) & $1.4 < r < 100$ &  $0.2 < r < 100$ \\
$|m_\mathrm{p\pi}-m_{\Lambda,\rm{PDG}}|$ (MeV/$c^2$) & $<10$ & $<7$ \\
\hline
$\dcar$ of pions decaying from $\Xi^{-}$ to the PV (cm) &  & $>0.015$ \\
DCA $\Xi^{-}$ daughters (cm) & & $<1.6$ \\
$\cos{\theta_\Xi}$ & & $>0.97$ \\
$r$($\Xi$) (cm) & & $0.2<r<100$ \\
$|m_\mathrm{\Lambda\pi}-m_{\Xi,\rm{PDG}}|$ (MeV/$c^2$) &  & $<7$ \\
\hline 
$|y|$ of reconstructed resonance & \multicolumn{2}{c}{$< 0.5$} \\

\hline
\end{tabular}
\end{table}

The $\Xi^{-}$ from $\Xi(1530)^{0}$ decays are selected by requiring that the accompanying pion called "bachelor" pion, $\pi^{-}_{\rm{Bach}}$, and the $\Lambda$ baryon stem from a common point in space by imposing that the DCA between $\Xi^{-}$ daughters is < 1.6 cm (see Fig.\ref{fig:decaymode}(b)) and further demanding the $\pi^{-}_{\rm{Bach}}$ from $\Xi^{-}$ to be a secondary particle, i.e. a track sufficiently apart from the primary vertex ($\mathrm{DCA}~\pi^{-}_{\rm{Bach}}$ to PV > 0.015 cm). The DCA between the $\lmb$ decay products, $\pi$ and p ($\rm{DCA}$~$\Lambda$~$\rm{daughters}$ in Fig.\ref{fig:decaymode}(b)) is required to be lower than 1.4 cm. The $\Lambda$ baryon itself is a secondary decay product in the decay topology, hence the $\rm{DCA}$ of $\Lambda$ to the PV ($\rm{DCA}$~$\Lambda$~$\rm{to}$~$\rm{PV}$ in Fig.\ref{fig:decaymode}(b)) is 
required to be larger than 0.07 cm. Selections on the invariant masses of the daughter particles, the cosine of the pointing angles ($\theta_\Lambda$ and $\theta_\Xi$), and the transverse distance from PV $r$($\Lambda$) and $r$($\Xi$) are applied to optimise the balance between purity and efficiency of each particle species. The selection criteria for $\sigmapm$ and $\xizero$ are listed in Table~\ref{tab:topological cut}.

Finally, all reconstructed resonances are required to be in the rapidity interval ($|y|<0.5$).

\subsection{Signal extraction}
\label{subsec:sigext}
\begin{figure}[!tb]
    \centering
        \begin{subfigure}{0.495\textwidth}
    \includegraphics[width=\linewidth]{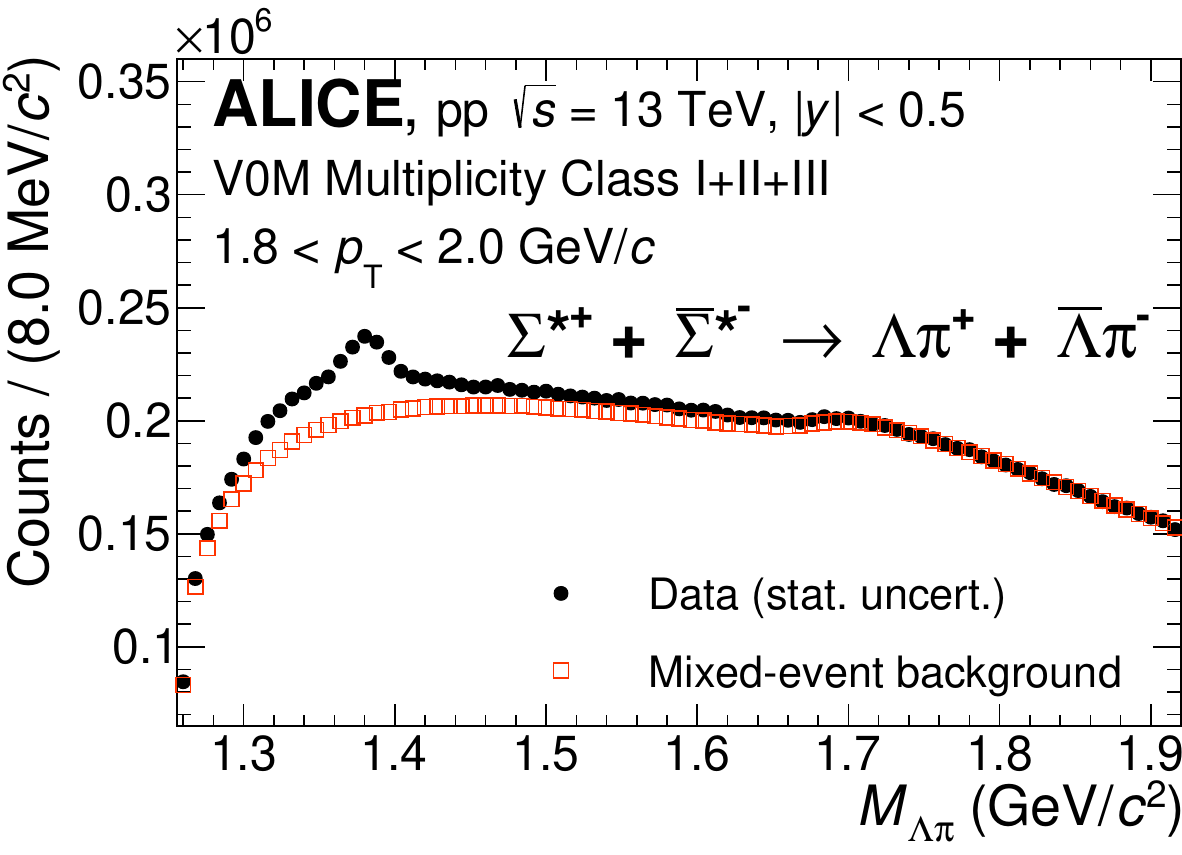}
    \caption{}
    \end{subfigure}
        \begin{subfigure}{0.495\textwidth}
    \includegraphics[width=\linewidth]{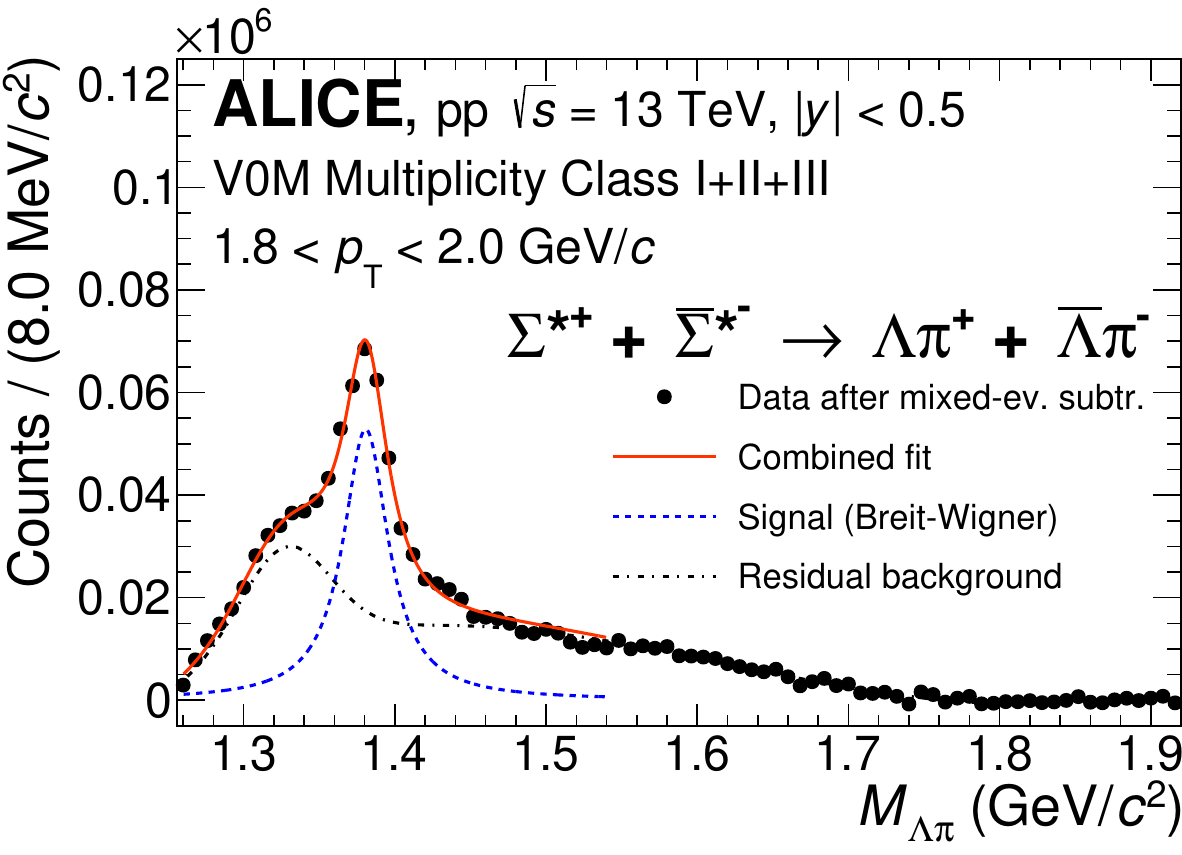}
    \caption{}
    \end{subfigure}
    \begin{subfigure}{0.495\textwidth}
    \includegraphics[width=\linewidth]{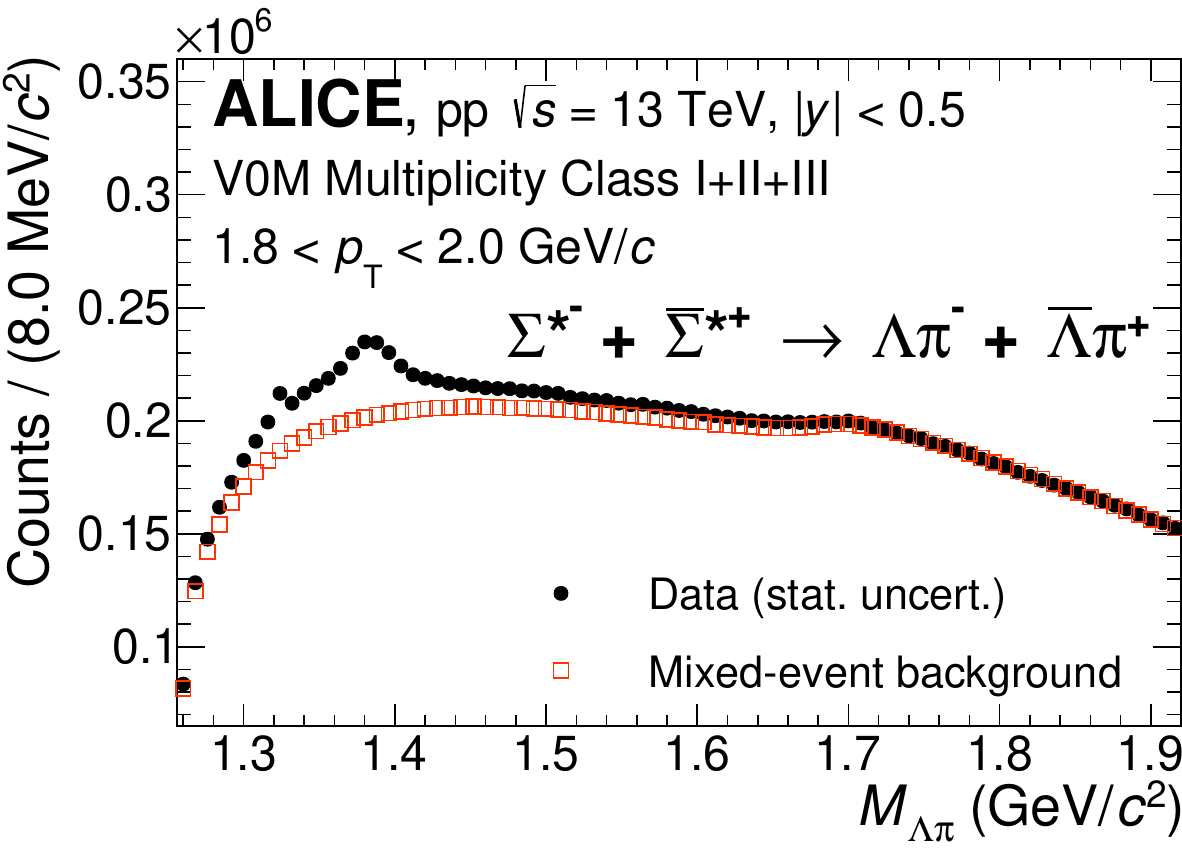}
    \caption{}
    \end{subfigure}
        \begin{subfigure}{0.495\textwidth}
    \includegraphics[width=\linewidth]{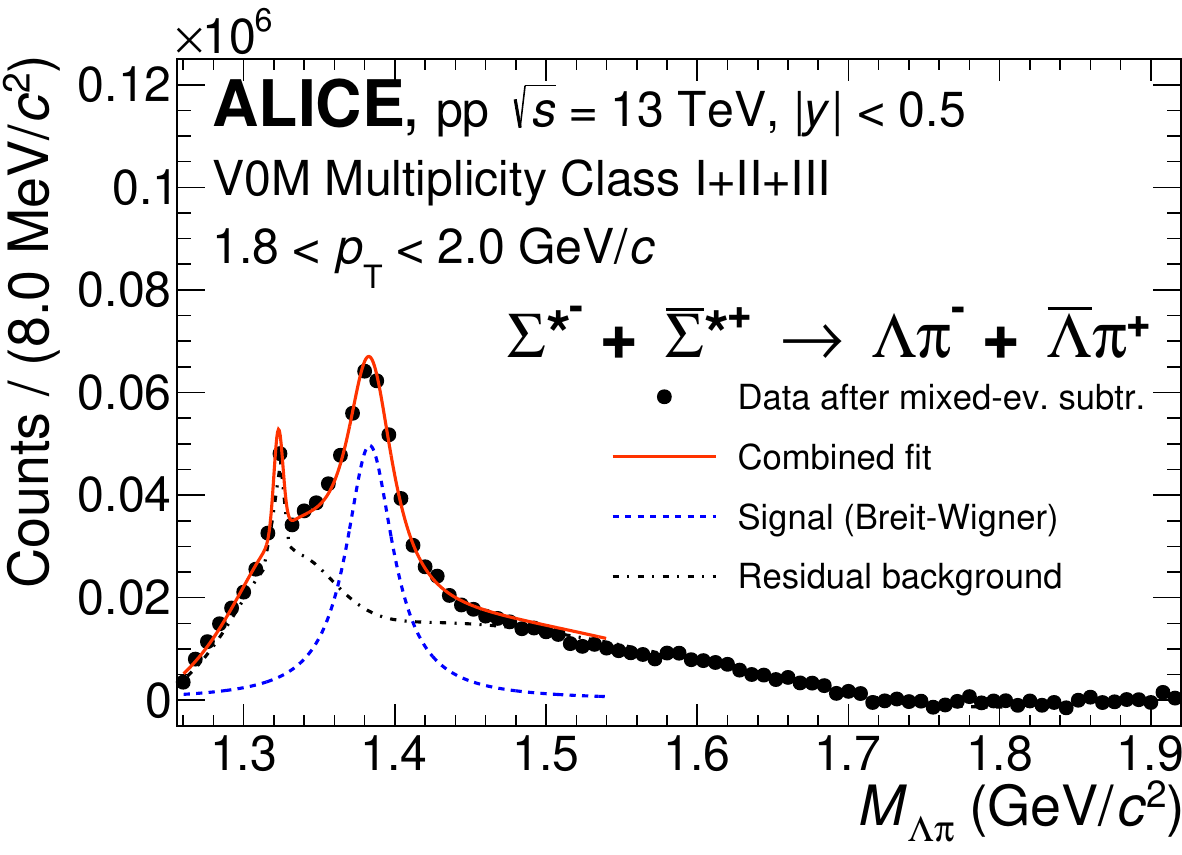}
    \caption{}
    \end{subfigure}
    \caption{The invariant mass distribution of $\Lambda\pi^{+} + \overline{\Lambda}\pi^{-}$ pairs (a) and the charge conjugates (c) in $|y|<0.5$ produced in \pp collisions at $\sqrt{s} = 13$ TeV for $1.8 <p_\mathrm{T,\,\Lambda\pi}< 2.0$ GeV/$c$ and the I+II+III multiplicity class (full black circles). The combinatorial background estimated with the event mixing technique is shown as open red squares in the (a) and (c) panels, whereas the invariant mass distributions after combinatorial background subtraction are shown in the (b) and (d) panels together with the fits to the signal and the residual background contributions. The solid red curves are the results of the combined fit and the dashed black lines represent the residual background.}
    \label{fig:sigextsigma}
\end{figure}

\begin{figure}[!htb]
    \centering
    \begin{subfigure}{0.495\textwidth}
        \includegraphics[width=\linewidth]{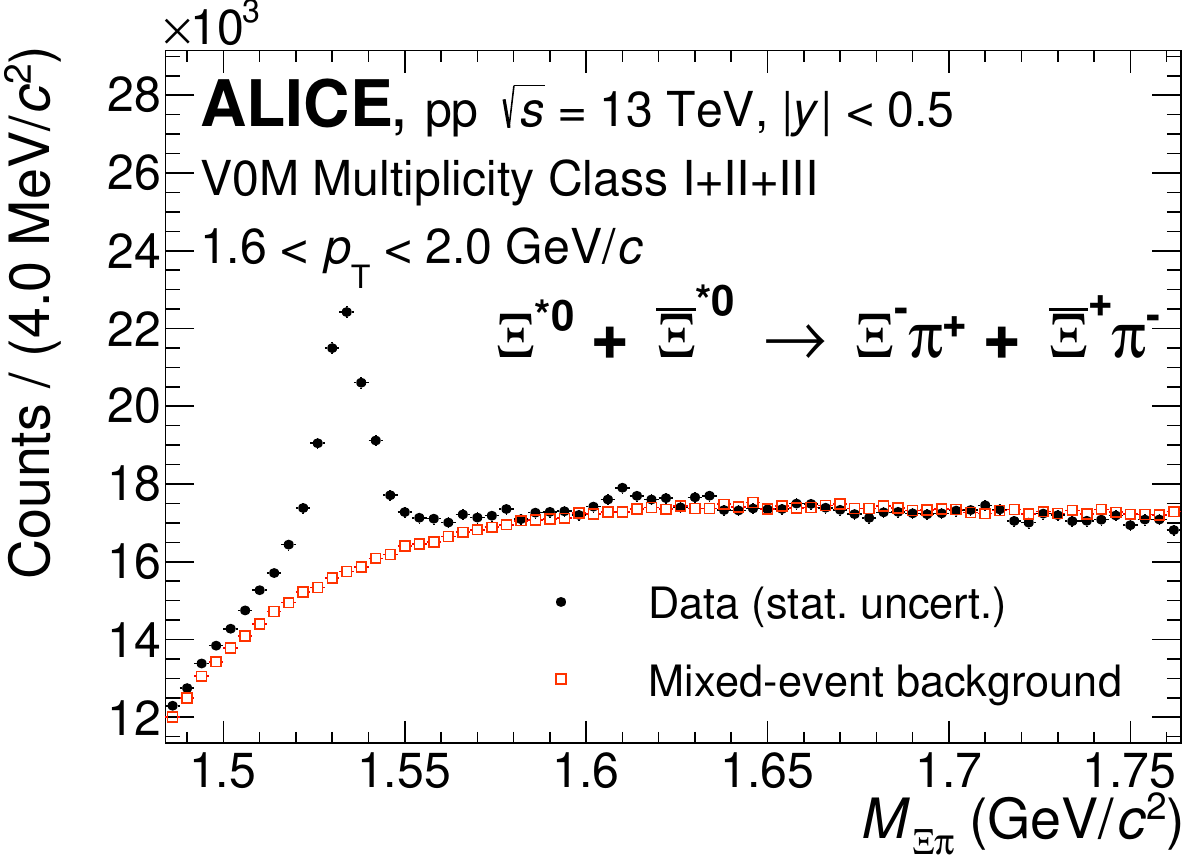}
        \vskip-0.5em
        \caption{}
    \end{subfigure}
    \begin{subfigure}{0.495\textwidth}
        \includegraphics[width=\linewidth]{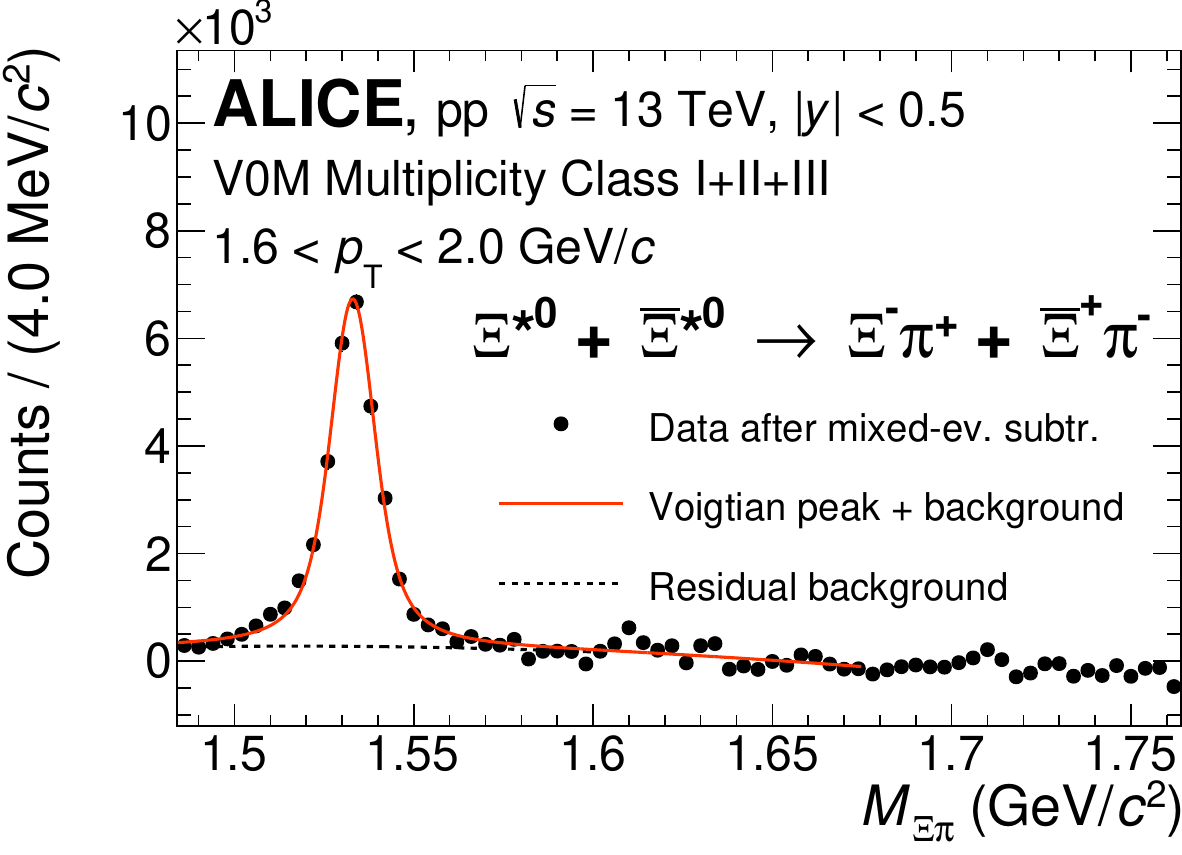}
        \vskip-0.5em
        \caption{}
    \end{subfigure}
    \caption{
    The invariant mass distribution of $\Xi^-\pi^+ + \Xi^+\pi^-$ pairs in $|y|<0.5$ produced in \pp collisions at $\sqrt{s} = 13$ TeV for $1.6 <p_\mathrm{T,\,\Lambda\pi}< 2.0$ GeV/$c$ and the I+II+III multiplicity class (full black circles). The combinatorial background estimated with the event mixing technique is shown as open red squares in panel (a), whereas the invariant mass distribution after combinatorial background subtraction is shown in panel (b) together with the fits to the signal and the residual background contributions. The solid red curve is the result of the combined fit and the dashed black line represents the residual background.}
    \label{fig:sigextxi}
\end{figure}

The selected hyperons and primary pions from the same event are combined into pairs to compute the invariant mass in given $p_\mathrm{T}$ intervals and in the region $|y|<0.5$. The invariant mass distributions of $\Lambda\pi^+$  ($\overline{\Lambda}\pi^{-}$) and $\Lambda\pi^-$ ($\overline{\Lambda}\pi^{+}$) pairs for $1.8 <p_\mathrm{T}< 2.0$ GeV/$c$ and $\Xi^-\pi^+$ ($\Xi^+\pi^-$) pairs for $1.6 <p_\mathrm{T}< 2.0$ GeV/$c$ are shown in Figs.~\ref{fig:sigextsigma} and~\ref{fig:sigextxi}, respectively. In order to increase the significance of the signal, the invariant mass distributions of $\sigmap$ and $\barsigmam$ are summed together in Fig.~\ref{fig:sigextsigma}a and~\ref{fig:sigextsigma}b. Similarly, the distributions for $\sigmam$ and $\barsigmap$ (Fig.~\ref{fig:sigextsigma}c and~\ref{fig:sigextsigma}d), as well as $\xizero$ and $\barxizero$, are also summed (Fig.~\ref{fig:sigextxi}). The combinatorial background distributions in the figures are estimated through an event-mixing technique where $\lmb\pi$ ($\Xi\pi$) pairs are formed by combining $\Lambda$ ($\Xi$) candidates from one event with $\pi$ from different events. Each event is combined with nine others. To minimise distortions due to the different positions of the PV and to ensure a similar event structure, the events entering the pool for mixing are requested to i) have a similar PV position in the $z$ direction ($|\Delta z_\mathrm{PV}|<1$ cm) and ii) belong to the same multiplicity class. 

The mixed-event background and the same-event distributions are normalised in an invariant mass interval where they are supposed to overlap, away from the signal peak. 
The normalisation regions are $1.7 < M_{\Lambda\pi} < 1.9$~GeV/$c^2$ and $1.65 < M_{\Xi\pi} < 1.75$~GeV/$c^2$ for $\Lambda\pi$ and $\Xi\pi$ invariant mass distributions, respectively. 

The mixed-event background is subtracted from the same-event invariant mass distribution producing the histograms for $\sigmap$ ($\barsigmam$),  $\sigmam$ ($\barsigmap$) and $\xizero$ ($\barxizero$) candidates, which are reported in the right panels of Figs.~\ref{fig:sigextsigma} and~\ref{fig:sigextxi}. 
These distributions are fitted with a combination of a non-relativistic Breit-Wigner function or a Voigtian function to describe the signal peak and a function to describe the residual background of correlated pairs, which is detailed subsequently below, that remain after the subtraction of the combinatorial background estimated with the mixed-event technique. The Breit-Wigner function is used to describe the signal peak of $\sigmapm$ and the Voigtian function, with the width of the Lorentzian part fixed at the PDG value, is used for $\xizero$. The Voigtian function is the convolution of a Breit-Wigner function to the line shape of the resonance and a Gaussian function to account for the detector resolution that for $\xizero$ is not negligible in comparison with the width of the resonance. The fitting ranges are $1.28 < M_{\Lambda\pi} < 1.54$~GeV/$c^2$ for $\Sigma(1385)$ and $1.47 < M_{\Xi\pi} < 1.65$~GeV/$c^2$ for $\xizero$, as in previous analysis at $\sqrt{s}$ = 7 TeV~\cite{ALICE:2014zxz}.

The residual background consists of $\Lambda\pi$ ($\Xi\pi$) pairs originating from the decays of other particles.
For $\sigmapm$, a template function is implemented for the residual background based on Monte Carlo simulations.
As explained in~\cite{ALICE:2014zxz} where the reader is redirected for details, the residual background is partly due to the decays of other particles which have $\Lambda\pi$ among the decay products and partly due to the dynamics of the collision that is not removed from the subtraction of the mixed-event background.
In Fig.~\ref{fig:sigextsigma}c and~\ref{fig:sigextsigma}d the peak from $\Xi^{-} \rightarrow \Lambda + \pi^{-}$ ($\overline{\Xi}^{+}$ $\rightarrow \overline{\Lambda} + \pi^{+}$) is visible: this physical background is accounted for by an additional component entering the residual background for $\Sigma(1385)^{-}$ ($\overline{\Sigma}(1385)^{+}$) decays. For $\xizero$, a second-order polynomial is used. 

In each $\pt$ interval, the raw yield of $\Sigma(1385)^{\pm}$ and $\xizero$ is obtained by integrating  the Breit-Wigner function and Voigtian function, respectively.

\subsection{Corrections and normalisation}
\label{subsec:corrections}
The raw yields of $\sigmapm$ and $\xizero$ resonances for each $\pt$ interval are corrected for the geometrical acceptance and the reconstruction efficiency of the detector, and normalised for the branching ratios of the considered decay channels.

\begin{figure}[!htb]
    \centering
    \includegraphics[width=0.6\linewidth]{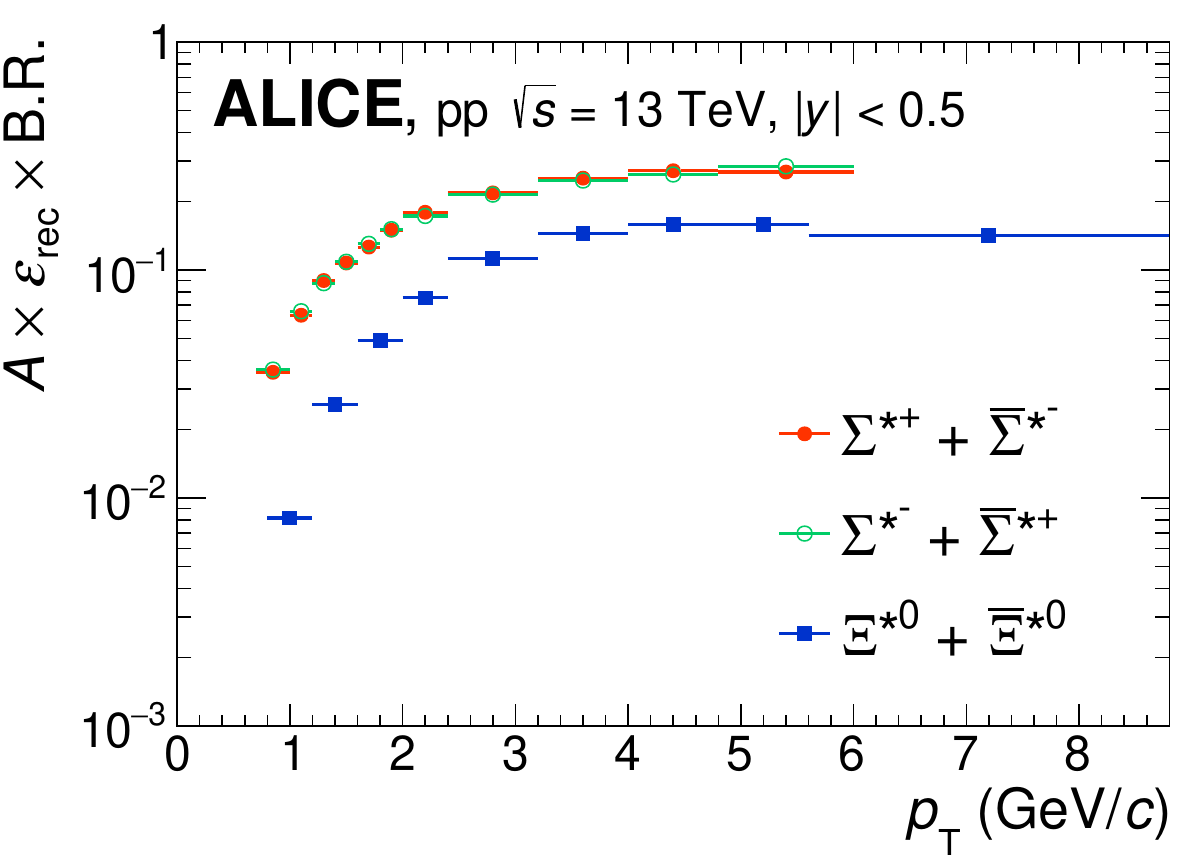}
    \caption{ The product of geometrical acceptance ($A$), reconstruction efficiency of the detector ($\epsilon_{\rm{rec}}$) and branching ratio (B.R.) for $\sigmapm$ and $\xizero$ resonances as a function of $\pt$ in $|y|<0.5$ obtained with simulations based on event generation with $\pythiaDef$ Monash 2013~\cite{Skands:2014pea} and particle transported with GEANT~3~\cite{Brun:1987ma}.   
    }
    \label{fig:receffi}
\end{figure}

The correction factors are estimated from Monte Carlo simulations based on the PYTHIA 8 event generator~\cite{Sjostrand:2007gs} with the Monash 2013 tune~\cite{Skands:2014pea} and on GEANT~3~\cite{Brun:1987ma} (v2-4-14) for the transport of particles through the ALICE detector. The $\sigmapm$ and $\xizero$ resonances from the simulation are reconstructed and selected by applying the same track quality, topological, and particle identification criteria as for the data. The generated and reconstructed resonances in $\pythiaDef$ are used to calculate the correction factors. The corrections, including acceptance, efficiency, and total branching ratio, $A\times\epsilon_{\rm{rec}}\times\rm{B.R.}$, for $\inelg$ events are shown in Fig.~\ref{fig:receffi}. Since no strong multiplicity dependence of the efficiency is observed as in Ref~\cite{ALICE:2019avo}, the values from the $\inelg$ events are used for the correction of the raw yields of all the multiplicity classes, and a $2\%$ systematic uncertainty, constant across $\pt$ bins, is assigned to account for residual differences.

As mentioned in section~\ref{sec:dataanalysis}, $\inelg$ selection is required. However, due to the inefficiency of the trigger, some $\inelg$ events are not selected and, in turn, all the $\sigmapm$ and $\xizero$ resonances produced in those events are not counted as well. A signal-loss correction, $f_{\rm{S.L.}}$, is thus applied, which accounts for resonances in non-triggered events. This is evaluated using the same simulations used to estimate the acceptance and efficiency. To calculate this correction factor, the simulated resonance $\pt$ spectrum before triggering and event selection is divided by the corresponding $\pt$ spectrum after those selections for each multiplicity class. This correction is more important for the lower multiplicity classes (10\% correction for class X, while only 1\% for class I+II+III).

Along with the correction of the yield, the number of events needs to be corrected for the inefficiencies of the trigger and the event selections, such as the primary vertex selection. The trigger efficiency $\epsilon_{\rm{Trig}}$ and the vertex selection efficiency $\epsilon_{\rm{Vertex}}$ are about 88\% and 98\%, respectively, for the lower-multiplicity classes and reach almost 100\% for the higher-multiplicity classes~\cite{ALICE:2019avo,ALICE:2018pal}. In the case of inelastic collisions, a global normalisation factor, $\epsilon_{\rm{Trig}}\times\epsilon_{\rm{Vertex}} = 0.74$ with a relative uncertainty of $2.5\%$~\cite{ALICE:2020jsh} is applied.

\section{Systematic uncertainties}

\begin{table}[!bt]
	\centering
	\caption{Summary of systematic uncertainties on the differential yield $\rm{d}^2\it{N}/\rm{d}\it{p}_{\rm{T}}\rm{d}\it{y}$ for the $\inelg$ event class. Negligible contributions are noted as negl.}
\label{table:systable}
\begin{tabular}{c|c|c|c}
Source of uncertainty               & $\sigmap$ & $\sigmam$ & $\xizero$ \\
 \hline \hline
$\pt$-dependent                        & \multicolumn{3}{l}{}          \\ \hline
Signal extraction                   & 4--10\%           & 8--15\%               & 1--17\%                      \\ 
Topological selection               & 1--10\%         & 1--10\%                & 1--3\%                      \\
TPC particle identification                               & 1--3\%        & 1--3\%                 & 1--3\%                      \\
Material budget                     & 0.8--8.3\%            & 0.8--8.3\%              & 1.2--5.1\%                      \\\hline
$\pt$-independent                      & \multicolumn{3}{l}{} \\ \hline
ITS--TPC matching                 & 3\%          & 3\%                & 3\%                      \\
Dependence of efficiency on multiplicity & 2\%          & 2\%        & 2\%                      \\
Branching Ratio & 1.1\%          & 1.1\%        & 0.3\%                      \\ 
Signal loss correction & negl. & negl.        & negl.                      \\ 
Trigger efficiency & negl. & negl.        & negl.                      \\ 
Vertex selection efficiency & negl.          & negl.        & negl.                      \\ \hline
Total                               & 6--17\%       & 9--20\%          & 4--19\%                     \\ \hline
\end{tabular}
\end{table}

Different sources of systematic uncertainty on the measured $\sigmapm$ and $\xizero$ observables ($\frac{\mathrm{d}^2N}{\mathrm{d}p_\mathrm{T}\mathrm{d}y}$, $\frac{\mathrm{d}N}{\mathrm{d}y}$, \meanpt) were considered:
the sources are essentially associated with the global tracking efficiency, track quality and topological selections, particle identification, signal extraction, and knowledge of the ALICE material budget. The systematic uncertainties are determined by varying the fit range and the selection criteria. The procedure applies to both $\pt$-dependent and $\pt$-independent uncertainties and is reconsidered for each $\pt$ interval and multiplicity class. The list of sources is given in Table 5 together with the range of values estimated for each of them.

The signal extraction uncertainty includes the uncertainty from the fitting procedure quantified by varying the fit range, and the normalisation range of the invariant mass distributions of the mixed-event background. The signal extraction is the main contribution to the total systematic uncertainty for $\sigmapm$, originating from the bump structure located on the left side of the signal peak (see Fig.~\ref{fig:sigextsigma} (b)). The signal extraction uncertainty on the $\sigmap$ is around 10\% in most of the $\pt$ and multiplicity intervals while the one of the $\sigmam$ is around 15\% due to the effect of the additional $\Xi^{-}$ peak in the background. The signal extraction uncertainty of the $\xizero$ is around 5--6 \% on average and reaches up to 17\%, but only in the case of the first $\pt$ bin in the highest multiplicity event (I+II+III) case caused by the poor significance of the signal. The systematic uncertainty originating from an imperfect description in the simulation of the variables utilised in the selection of displaced decay topologies are estimated by varying the criteria on the DCA and the cosine of the pointing angle of the particle decay products $\Lambda$ and $\Xi^{-}$, and their mass windows. Variations are performed around the nominal values of the selection variables one at a time, while fixing all the other ones. 
$\sigmapm$ show a relatively higher uncertainty for the topological selection than the one of $\xizero$, given that the $\Sigma(1385)$ family is indeed further affected by the residual background (see Fig.~\ref{fig:sigextsigma}).
The systematic uncertainty associated with particle identification is determined by using a tighter and a looser requirement on the TPC d$E$/d$x$. The uncertainty from the material budget of the ALICE detector is taken from the studies performed for the reconstruction of $\Lambda$ and $\Xi^{-}$ hyperons reported in~\cite{ALICE:2020jsh}. 
Finally, the systematic uncertainty on ITS--TPC matching efficiency of the first emitted pion from the resonances ($\pi_{\rm{first}}$ in Fig.~\ref{fig:decaymode}) was defined from the difference in the prolongation probability of TPC tracks to ITS points between data and Monte Carlo simulations. The matching efficiency uncertainty and the tiny variation of the reconstruction efficiency in different multiplicity classes are considered as fully $\pt$-correlated uncertainties. 
Other uncertainties such as the ones on signal loss correction, trigger efficiency, and vertex selection efficiency have a negligible contribution to this study. 

The total uncertainty is calculated as the quadratic sum of the uncertainties from the different sources. 
The $\pt$-independent uncertainties are considered but as a separate group for $\frac{\mathrm{d}^2N}{\mathrm{d}p_\mathrm{T}\mathrm{d}y}$ spectra and \meanpt quantities, since they do not affect the $\pt$ shape of the spectra but just their overall magnitude. On the other hand, such uncertainties are of special concern for $\pt$-integrated quantities such as $\frac{\mathrm{d}N}{\mathrm{d}y}$.
For quantities given along \dNchdeta, a special investigation was conducted to quantify the level of correlation of all systematic uncertainties along event multiplicity.

\section{Results}
\begin{figure}[hp]
    \centering
    \begin{subfigure}{0.485\textwidth}
    \includegraphics[width=\linewidth]{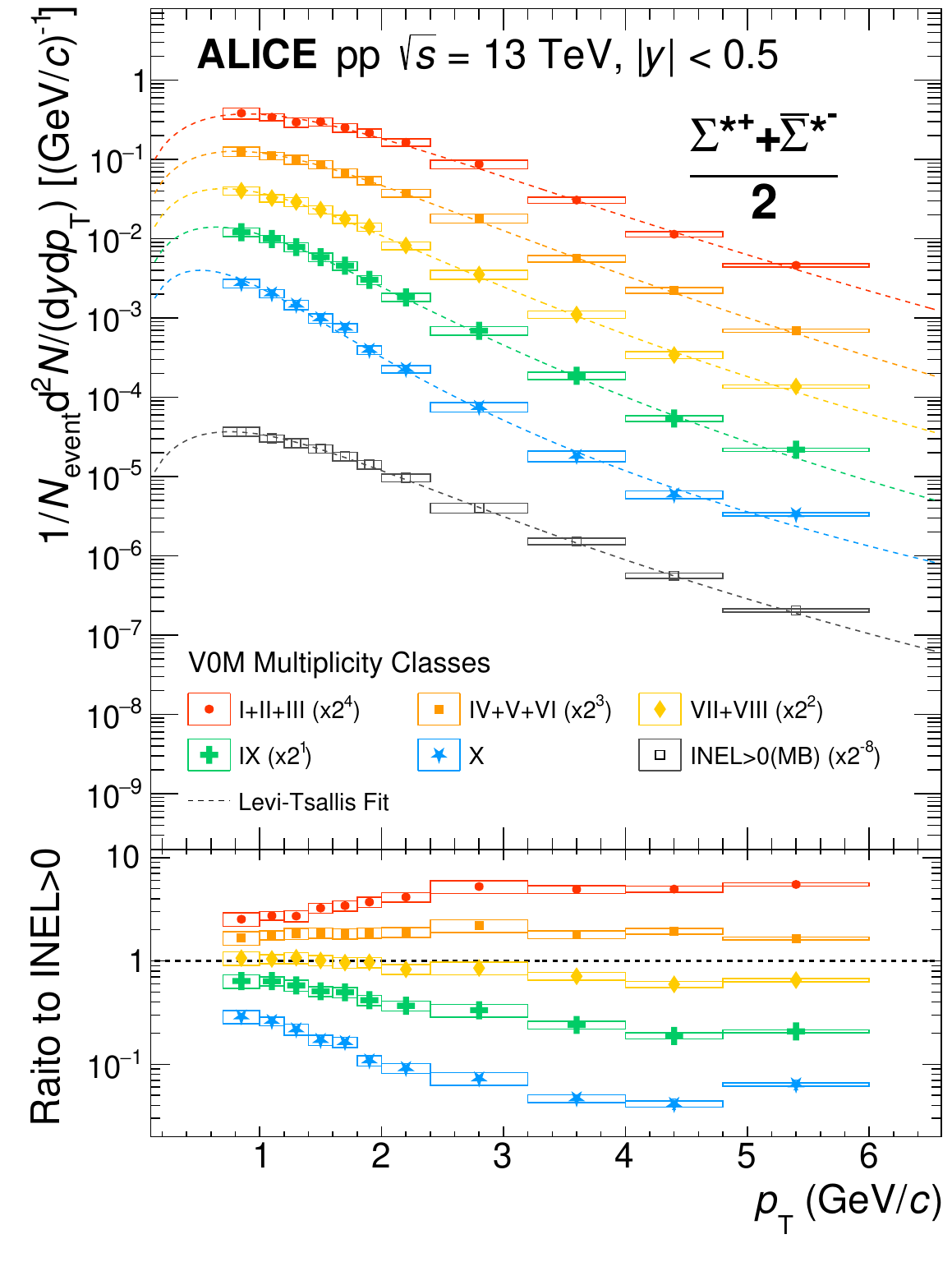}
    \vskip-0.5em
    \caption{}
    \end{subfigure}
    \begin{subfigure}{0.485\textwidth}
    \includegraphics[width=\linewidth]{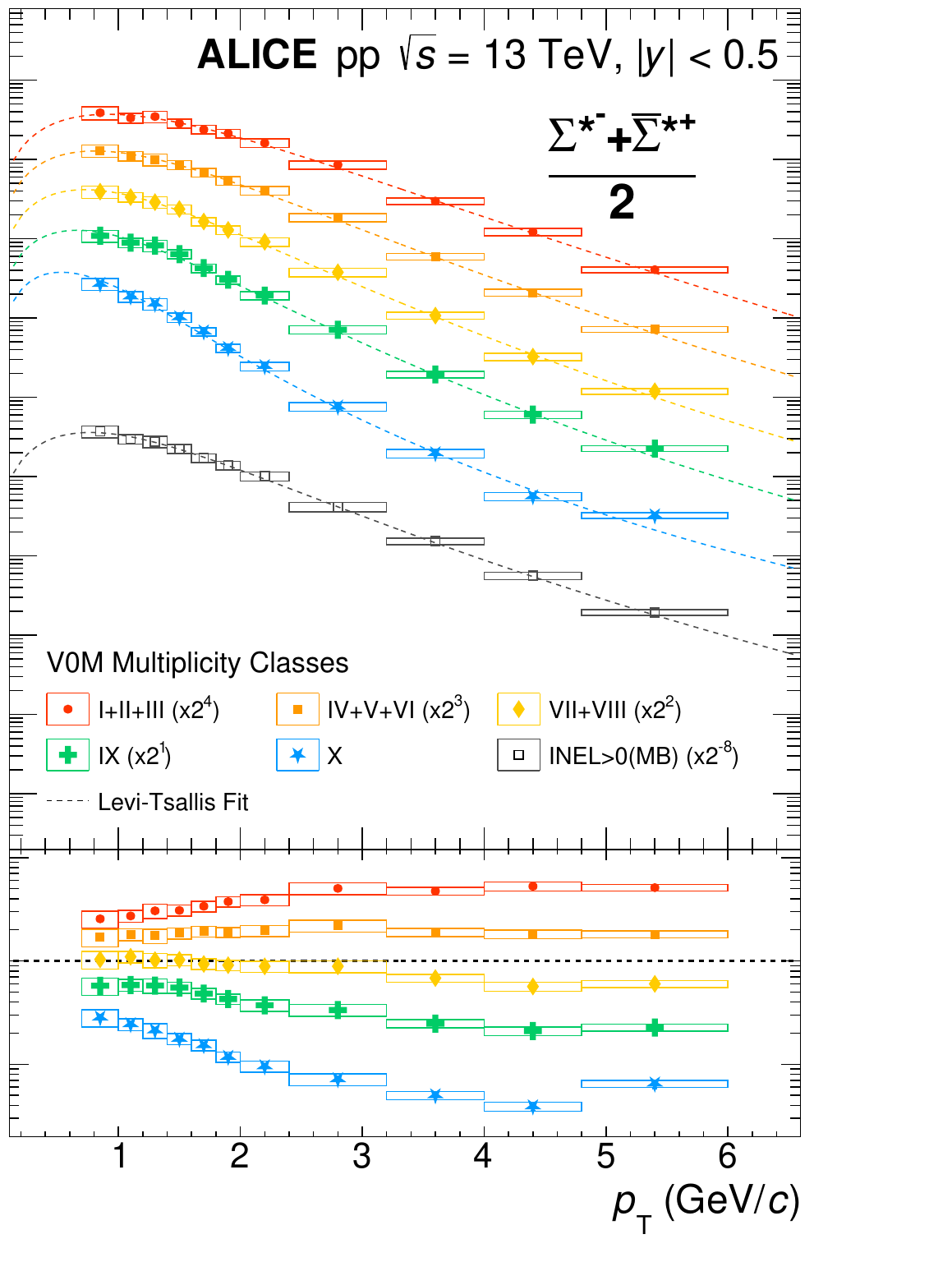}
    \vskip-0.5em
    \caption{}
    \end{subfigure}
    \begin{subfigure}{0.485\textwidth}
    \includegraphics[width=\linewidth]{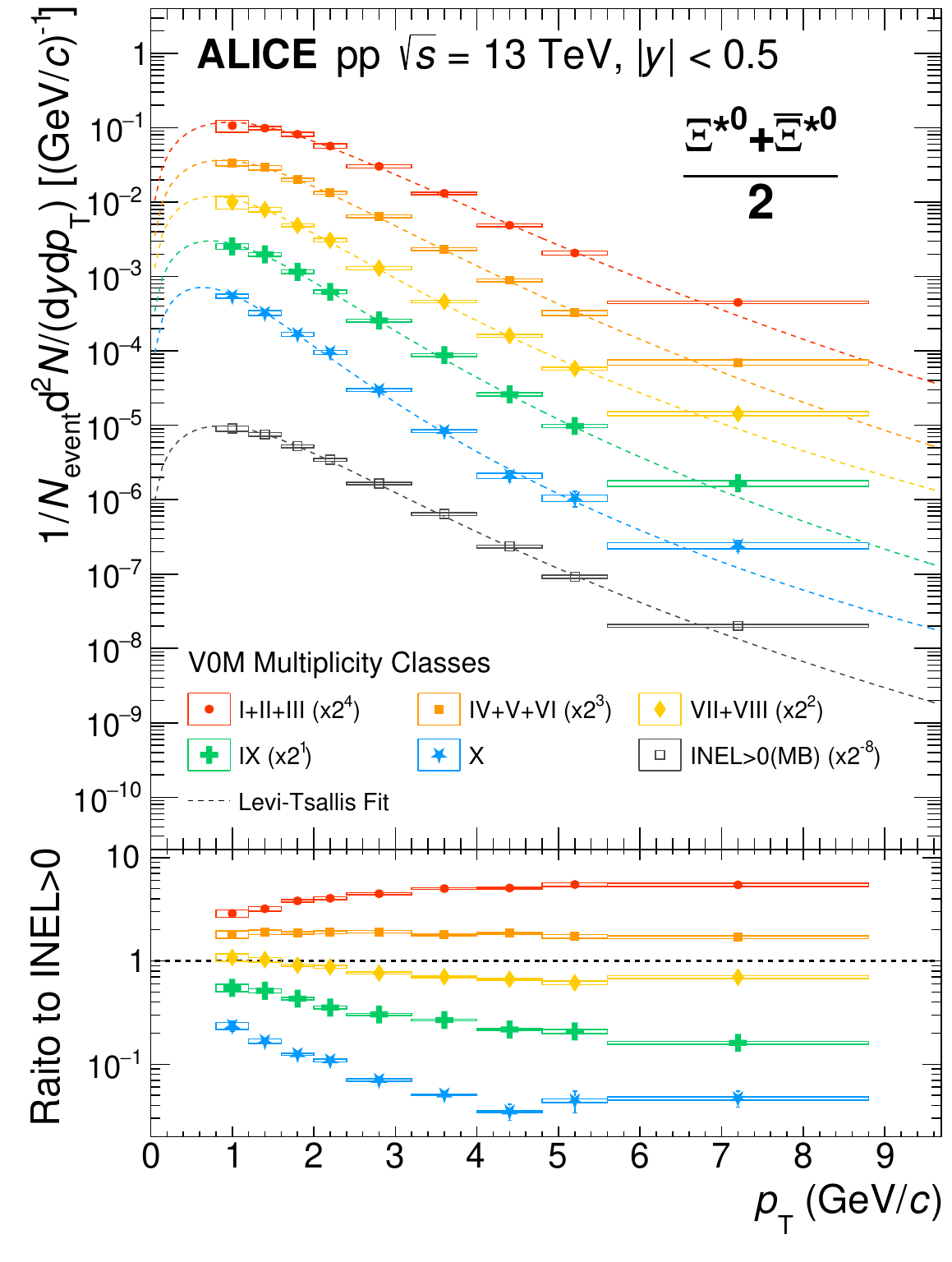}
    \vskip-0.5em
    \caption{}
    \end{subfigure}
    \caption{Transverse momentum spectra of $\sigmap$ (a), $\sigmam$ (b) and $\xizero$ (c) in pp collisions at $\cms$ = 13 TeV in multiplicity classes and for the inclusive case ($\inelg$). Statistical and total systematic uncertainties are shown by error bars and boxes, respectively. The bottom panels show the ratios of the multiplicity-dependent spectra to the $\inelg$ distributions. The systematic uncertainties on the ratios are obtained by considering only contributions of multiplicity-uncorrelated uncertainties described in Table~\ref{table:systable}. The dashed lines represent the fits to the spectra with the \levi function.}
    \label{fig:SigmaSigmaspectra}
\end{figure}

The $\pt$-differential yields ($\pt$ spectra)  for $\sigmap$, $\sigmam$, and $\xizero$ (and their antiparticles) in the various multiplicity classes, as well as the ratios of these spectra to the inclusive INEL$>0$ spectrum, are shown in Fig.~\ref{fig:SigmaSigmaspectra}. The $\pt$ spectra of $\sigmap$ and $\sigmam$ are identical within uncertainties.

For $\pt<4$ \GeVc, a hardening of the $\pt$ spectra from low to high-multiplicity events is clearly visible, while at higher $\pt$ the spectra have the same shape regardless of the multiplicity class, indicating that the processes that affect the shape of the $\pt$ spectra depending on the multiplicity of particles produced in the collision are dominant at low $\pt$. A similar behaviour was reported for other species, in collisions at the same energy~\cite{Acharya:2020zji}.

\begin{figure}[!htb]
    \centering
    \begin{subfigure}{0.49\textwidth}
    \includegraphics[width=\linewidth]{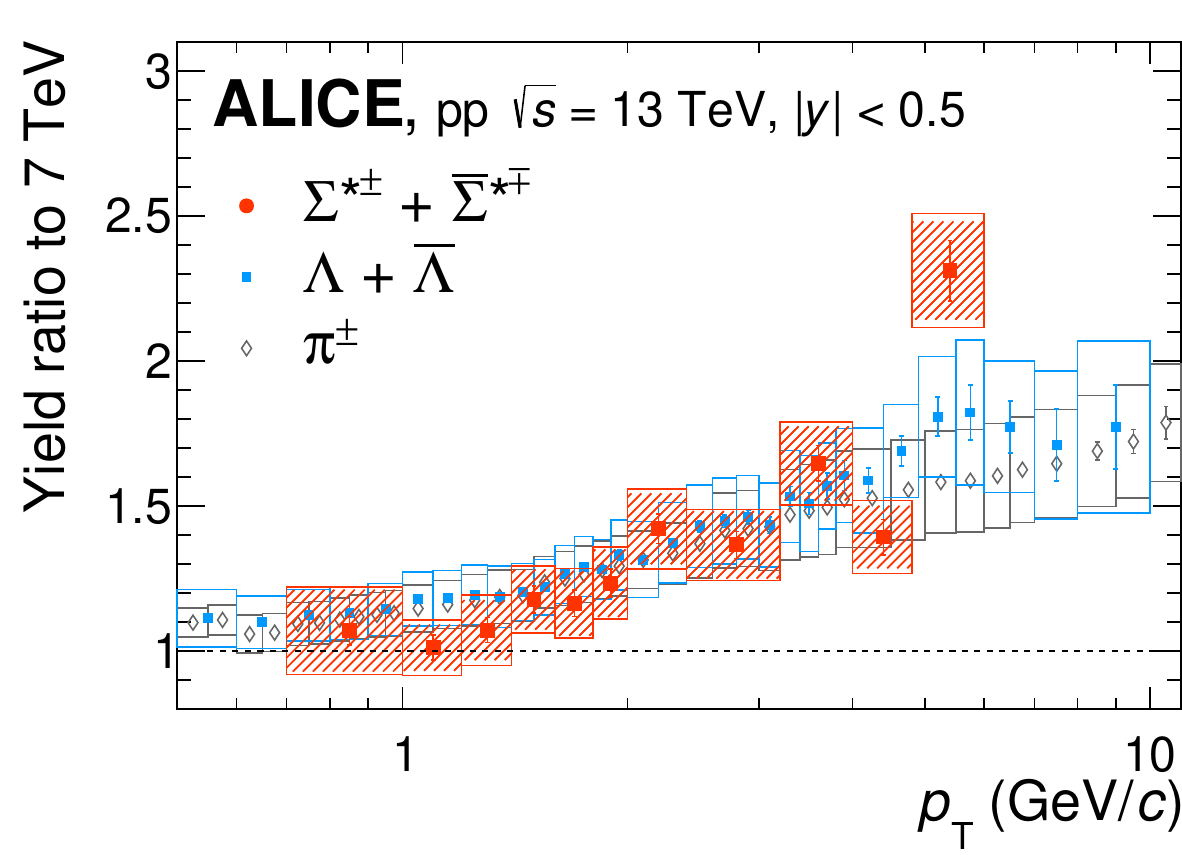}
    \vskip-0.5em
    \caption{}
    \end{subfigure}
    \begin{subfigure}{0.49\textwidth}
    \includegraphics[width=\linewidth]{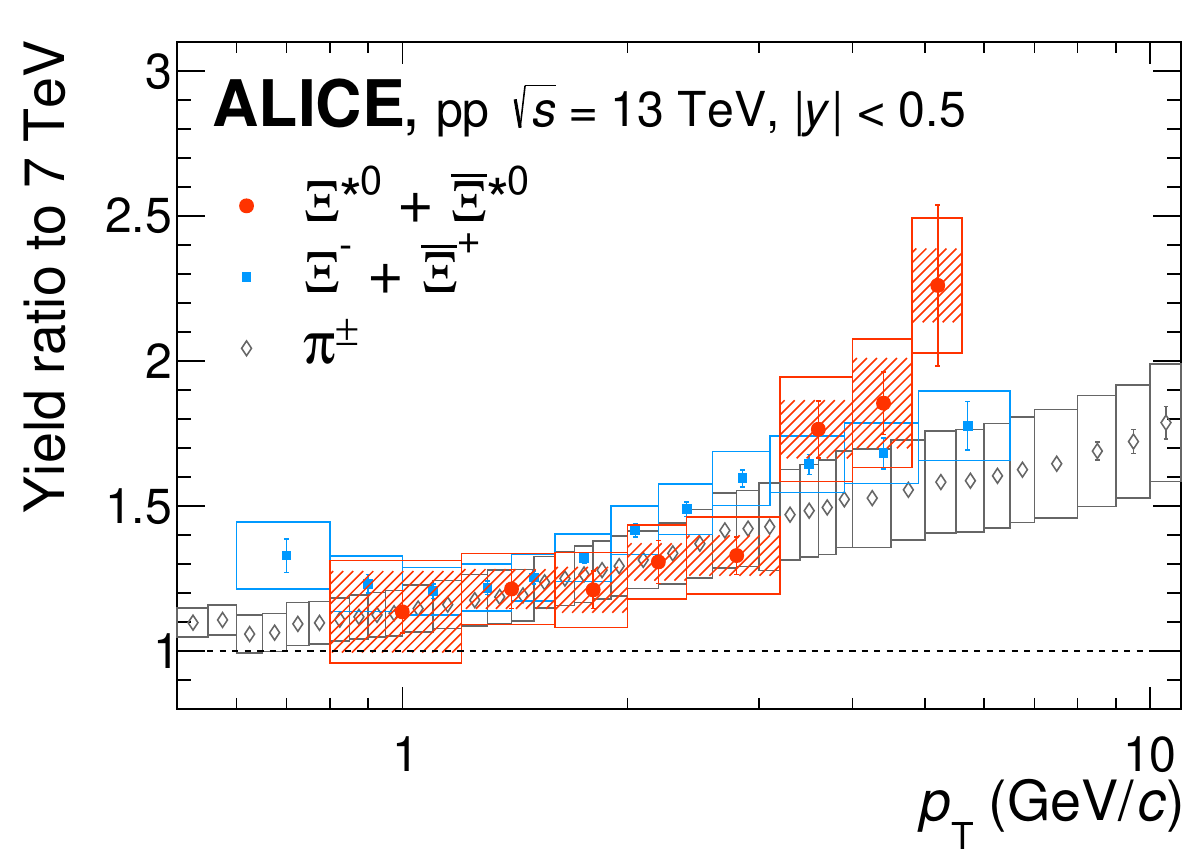}
    \vskip-0.5em
    \caption{}
    \end{subfigure}
    \caption{Ratios of transverse momentum spectra of $\sigmapm$ (a) and $\xizero$ (b) in inelastic \pp collisions at $\cms$ = 13 TeV to the ones in inelastic \pp collisions $\cms$ = 7 TeV~\cite{ALICE:2014zxz} compared with those of $\Xi^{-}$, $\Lambda$ and $\pi^{\pm}$~\cite{ALICE:2012yqk,ALICE:2020jsh}. The statistical and systematic uncertainties are shown as vertical error bars and boxes, respectively. In the present measurement, the shaded boxes represent the multiplicity-uncorrelated uncertainties.}     
    \label{fig:pTratio}
\end{figure}

Figure~\ref{fig:pTratio} shows the ratios of the transverse momentum spectra for inelastic \pp collisions at $\cms~=~13$~TeV to those at $\cms~=~7$~TeV  for $\sigmapm$ (left) and $\xizero$  (right). 
For both $\sigmapm$ and $\xizero$, the yield ratios at low $\pt$ ($\pt<2$ \GeVc) are slightly larger than unity, even though they are compatible with unity within the systematic uncertainties. The yield ratios are also consistent with being independent of $\pt$ at low $\pt$. These considerations suggest that the production mechanism of these resonances in the soft scattering regime is only mildly dependent on the collision energy in the measured energy range. 
For $\pt>2$ \GeVc, the ratios are observed to depart from unity, indicating a hardening of the $\pt$ spectra at $\cms~=~13$ TeV as compared with $\cms~=~7$ TeV. A similar behaviour was observed for the yield ratios of $\pi^{\pm}$, $\Lambda$ and $\Xi^{-}$~\cite{ALICE:2020jsh} which are shown in Figure~\ref{fig:pTratio} overlaid to the  $\sigmapm$ and $\xizero$ results.
The slope of the $\xizero$ yield ratios in Figure~\ref{fig:pTratio} (b) is compatible within uncertainties with the slope of $\Xi^{\mp}$ yield ratios and the rapid increase of $\xizero$ for $\pt>3$ \GeVc needs to be confirmed in the higher $\pt$ region with more precise measurements on larger data samples for further interpretations on any distinction.

To calculate the total yields of $\sigmapm$ and $\xizero$ integrated over $\pt$ (d$N$/d$y$) and their mean transverse momentum \meanpt, the measured $\pt$-differential spectra are fitted with a \levi function~\cite{Tsallis:1987eu} defined as:

\begin{equation}
  \frac{1}{N_\mathrm{event}} \frac{\mathrm{d}^2N}{\mathrm{d}p_\mathrm{T}\mathrm{d}y} = p_\mathrm{T} N' \frac{(n-1)(n-2)}{nC\left[nC+m_0(n-2)\right]} \left[ 1+\frac{\sqrt{p^{2}_{\rm{T}}+m^{2}_{0}}-m_{0}}{nC} \right]^{-n},
\end{equation}

where $N_{\rm{event}}$ is the number of events in a given multiplicity class, $m_{0}$ is the world-average mass of the particle, and $n$, $C$, and the integrated yield
$N'$ are free parameters of the fit. This function is successfully used to describe most of the identified particle spectra in pp collisions~\cite{ALICE:2019avo,ALICE:2020jsh}. The \levi functions obtained by fitting the $\pt$ spectra in the different multiplicity classes are shown as dashed lines in Fig.~\ref{fig:SigmaSigmaspectra}. 

The value of d$N$/d$y$ is obtained by integrating the measured spectrum from $\pt$ = 0.7 (0.8) to 6.0 (8.8) \GeVc for $\sigmapm$ ($\xizero$) and the extrapolated fitting curve in the unmeasured regions down to $\pt=0$ and up to $\pt=10$ \GeVc. 
The \meanpt is defined as $\frac{\sum_{j}(\overline{\it{p}_{T,j}} \times \rm{d}\it{p}_{T,j} \times I_{j})}{\rm{d}\it{N} / \rm{d}\it{y}}$, where $j$ means each $\pt$ bin, $\overline{\it{p}_{T,j}}$ means bin center, $\rm{d}\it{p}_{T,j}$  mean bin width and $I_{j}$ means measured $\pt$-differential yield.
Similar to the d$N$/d$y$, \meanpt is computed using the measured spectra in the  $\pt$ interval of the measurement and the \levi function outside this range. The values of d$N$/d$y$ and \meanpt are reported in Table~\ref{tab:dNdetaAndMeanpT}.

The fractions of extrapolated particle yield at low $\pt$ for different multiplicity classes are 26--43\% and 21--43\% depending on the multiplicity class for $\sigmapm$ and $\xizero$, respectively. The extrapolated yields in the high-$\pt$ region are negligible. Alternative functions such as the Boltzmann, Fermi--Dirac, $m_{\rm{T}}$-exponential, $\pt$-exponential, and blast-wave functions are employed to estimate the systematic uncertainty of this extrapolation, which amounts to about 3--4\% for both resonances. The fit functions used for the systematic study are listed in Appendix~\ref{sec:fitfunctions}. As reported in Table~\ref{tab:dNdetaAndMeanpT}, the relative uncertainties of the INEL results are larger than those of the $\inelg$ results due to the propagation of the uncertainty on the normalisations, $\epsilon_{\rm{Trig}}$, $\epsilon_{\rm{Vertex}}$, and $f_{\rm{S.L.}}$ which are slightly different for the INEL and the $\inelg$ samples.

\begin{table}[h]
\centering
\caption{The values of d$N$/d$y$ and \meanpt for multiplicity-integrated spectra (INEL, $\inelg$) and for each multiplicity class. Statistical (first one), total systematic (second one) and multiplicity-uncorrelated systematic (third one, in brackets) uncertainties are quoted. The multiplicity-uncorrelated systematic uncertainties are not an additional source here but must be considered as a component of the total systematic uncertainties.}
\label{tab:dNdetaAndMeanpT}
\begin{tabular}{@{}c|c|c|c@{}}
\toprule
Baryon & Event Class & d$N$/d$y$ ($\times 10^{-3}$) & \meanpt (\GeVc) \\ \midrule
\multirow{7}{*}{$\xizero$} & $\inelg$ & $4.29\pm0.05\pm0.20\,(0.16$) & $1.44\pm0.01\pm0.06\,(0.05$) \\ \cmidrule(l){2-4}
       & I+II+III & $14.29\pm0.15\pm1.07\,(0.78)$ & $1.64\pm0.01\pm0.07\,(0.06)$ \\ 
       & IV+V+VI & $7.96\pm0.10\pm0.57\,(0.39)$ & $1.44\pm0.01\pm0.06\,(0.06)$ \\ 
       & VII+VIII & $4.42\pm0.07\pm0.47\,(0.41)$ & $1.30\pm0.01\pm0.07\,(0.07)$ \\ 
       & IX & $2.14\pm0.05\pm0.18\,(0.14)$ & $1.22\pm0.01\pm0.05\,(0.05)$ \\ 
       & X & $0.85\pm0.05\pm0.09\,(0.08)$ & $1.05\pm0.03\pm0.05\,(0.04)$ \\  \cmidrule(l){2-4}
       & INEL & $2.98\pm\num{0.07}\pm\num{0.15}\,(\num{0.11})$ & $1.45\pm\num{0.02}\pm\num{0.06}\,(\num{0.05})$ \\  \midrule
\multirow{8}{*}{$\sigmapm$} & $\inelg$ & $14.50\pm0.11\pm1.33\,(0.82)$ & $1.29\pm0.01\pm0.05\,(0.03)$ \\ \cmidrule(l){2-4}
       & I+II+III & $45.02\pm0.72\pm5.26\,(3.24)$ & $1.50\pm0.01\pm0.11\,(0.11)$ \\ 
       & IV+V+VI & $26.53\pm0.50\pm2.97\,(2.06)$ & $1.33\pm0.02\pm0.07\,(0.07)$ \\ 
       & VII+VIII & $15.19\pm0.36\pm1.89\,(1.37)$ & $1.19\pm0.02\pm0.07\,(0.06)$ \\ 
       & IX & $8.39\pm0.25\pm1.22\,(0.77)$ & $1.09\pm0.02\pm0.06\,(0.05)$ \\ 
       & X & $3.66\pm0.14\pm0.70\,(0.46)$ & $0.92\pm0.02\pm0.06\,(0.06)$ \\ \cmidrule(l){2-4} 
       & INEL & $10.91\pm\num{0.18}\pm\num{0.89}\,(\num{0.59})$ & $1.27\pm\num{0.01}\pm\num{0.04}\,(\num{0.02})$ \\ \bottomrule
\end{tabular}
\end{table}

\begin{figure}[!tb]
    \centering
    \includegraphics[width=\linewidth]{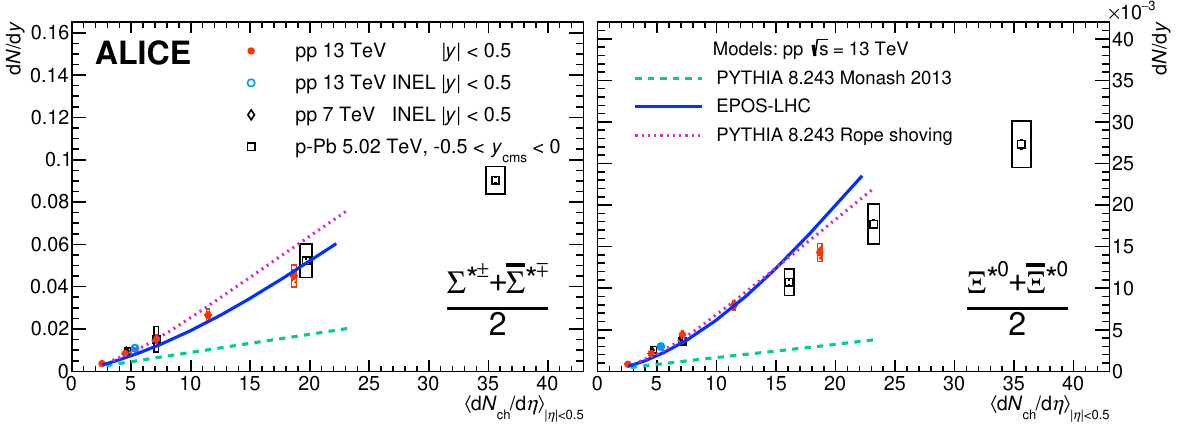}
    \caption{The $\pt$-integrated yields as a function of charged-particle pseudorapidity density $\avdndeta_{|\eta|<0.5}$ for $\sigmapm$ (left) and $\xizero$ (right) compared with the measurements in \pp collisions at \seven~\cite{ALICE:2014zxz} and \pPb collisions at \fivenn~\cite{ALICE:2017pgw}. The open and shaded boxes represent the total and multiplicity-uncorrelated systematic uncertainties, respectively. The measured points are compared with predictions from different event generators, namely EPOS-LHC~\cite{Pierog:2013ria}, $\pythiaDef$ with Monash 2013 tuning~\cite{Skands:2014pea}, and $\pythiaDef$ with Rope shoving ~\cite{Bierlich:2014xba,Bierlich:2015rha,Bierlich:2016vgw,Bierlich:2017vhg}. (Appendix~\ref{sec:ropetune}). The predictions are obtained for pp collisions at \thirteen on $\inelg$ events.}
    \label{fig:yield}
\end{figure}

\begin{figure}[!htb]
    \centering
    \includegraphics[width=\linewidth]{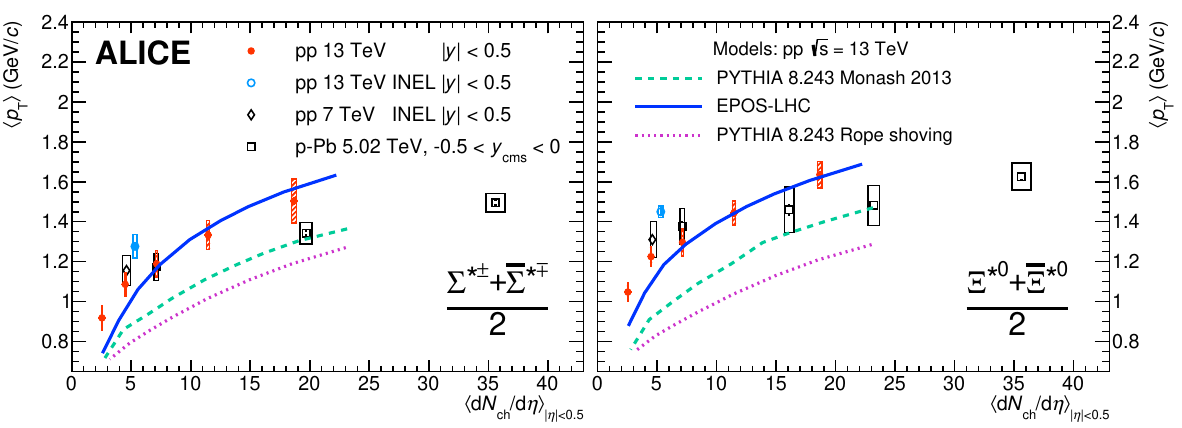}
    \caption{The \meanpt as a function of charged-particle pseudorapidity density for $\sigmapm$ (left) and $\xizero$ (right) compared with the measurements in \pp collisions at \seven~\cite{ALICE:2014zxz} and \pPb collisions at \fivenn~\cite{ALICE:2017pgw}. The open and shaded boxes represent the total and multiplicity-uncorrelated systematic uncertainties, respectively. The measured points are compared with predictions from different event generators, namely EPOS-LHC~\cite{Pierog:2013ria}, $\pythiaDef$ with Monash 2013 tuning~\cite{Skands:2014pea}, and $\pythiaDef$ with Rope shoving ~\cite{Bierlich:2014xba,Bierlich:2015rha,Bierlich:2016vgw,Bierlich:2017vhg}. (Appendix~\ref{sec:ropetune}). The predictions are obtained for pp collisions at \thirteen, based on $\inelg$ events.}
    \label{fig:mean}
\end{figure}

Figure~\ref{fig:yield} shows the $\pt$-integrated yields, d$N$/d$y$, as a function of the charged-particle pseudorapidity density at midrapidity and in comparison with the values previously measured in pp collisions at $\cms = 7$ TeV~\cite{ALICE:2014zxz} and in \pPb collisions at $\snn = 5.02$ TeV~\cite{ALICE:2017pgw} by the ALICE Collaboration. The comparison suggests that the integrated yields depend only on the multiplicity, irrespective of the collision energy and system. This is consistent with previous results at the LHC~\cite{ALICE:2019avo,Acharya:2020zji}, highlighting the fact that the mechanisms of particle production are related essentially to the event properties that determine the multiplicity. The measurements are compared with the predictions of different event generators, namely EPOS-LHC~\cite{Pierog:2013ria}, $\pythiaDef$ with Monash 2013 tuning~\cite{Skands:2014pea}, and $\pythiaDef$ with Rope shoving ~\cite{Bierlich:2014xba,Bierlich:2015rha,Bierlich:2016vgw,Bierlich:2017vhg}. Both EPOS-LHC and PYTHIA8 are QCD-inspired event generators. The EPOS-LHC includes a modelling of the collective behaviour implemented via a core-corona approach~\cite{Werner:2007bf}. Instead, $\pythiaDef$ with Monash 2013 does not include a collective expansion and is based on the Lund string fragmentation model. $\pythiaDef$ with Rope shoving describes multiparton interactions by allowing nearby strings to shove each other and form a colour "rope" from overlapping strings.
The model predictions are reported for pp collisions at $\cms = 13$ TeV. The increasing trend of the \dndyres of $\sigmapm$ and $\xizero$ with increasing \avdndeta for pp collisions is qualitatively reproduced by the different event generator models. However, $\pythiaDef$ with Monash 2013 tuning (green dashed curve) predicts a much milder increase for both resonances, whereas EPOS-LHC (blue solid curve) and $\pythiaDef$ with Rope shoving (purple dotted curve, see details of the tune settings in Appendix~\ref{sec:ropetune}) describe the measured \dndyres of both resonances within the uncertainties, with possibly a small tension between the EPOS-LHC prediction and the $\xizero$ data in the highest multiplicity interval.

Figure~\ref{fig:mean} shows the mean transverse momentum \meanpt for \sigmapm and \xizero as a function of the charged-particle pseudorapidity density. The values obtained in pp collisions at $\cms = 7$ TeV~\cite{ALICE:2014zxz} and in \pPb collisions at $\snn = 5.02$ TeV~\cite{ALICE:2017pgw} are reported in the same figure. The increasing trend of the \meanpt as a function of multiplicity in \pp collisions is steeper than the one in \pPb collisions for both $\sigmapm$ and $\xizero$, consistent with what is observed for unidentified charged particles in \pp collisions at $\cms$~=~7~TeV and \pPb collisions at $\snn$ = 5.02 TeV~\cite{Abelev:2013bla}. 
The increasing trend and the measured values are well described by EPOS-LHC.
Both $\pythiaDef$ with Monash 2013 tuning and $\pythiaDef$ with Rope shoving predict an increasing trend, however, they underestimate the data in the full range of the pp measurements.

\begin{figure}[h]
    \centering
    \includegraphics[width=\linewidth]{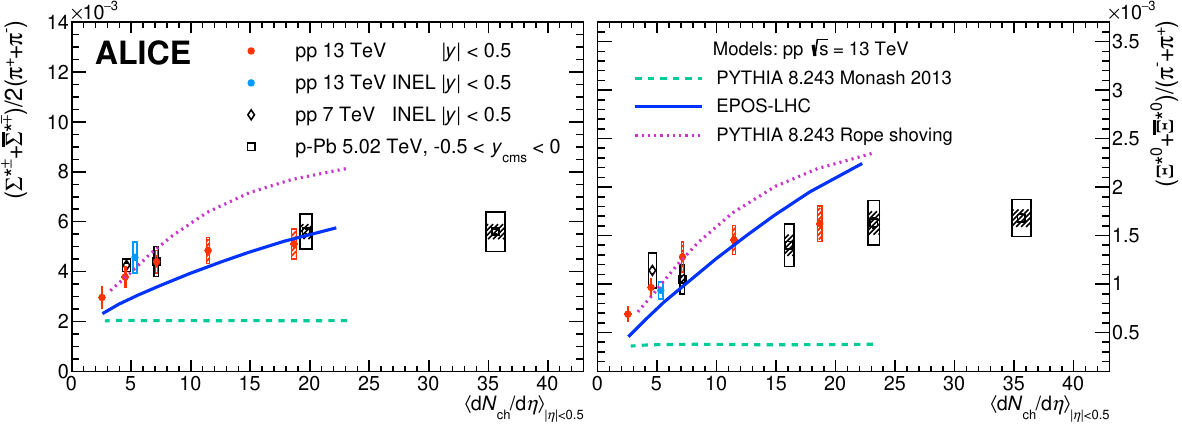}
    \caption{Ratio of the resonance to pion $\pt$-integrated  yield as a function of the charged-particle pseudorapidity density for $\sigmapm$ (left) and $\xizero$ (right). The open and shaded boxes represent the total and multiplicity-uncorrelated systematic uncertainties, respectively. The measured points are compared with predictions from different event generators, namely EPOS-LHC~\cite{Pierog:2013ria}, $\pythiaDef$ with Monash 2013 tuning~\cite{Skands:2014pea}, and $\pythiaDef$ with Rope shoving ~\cite{Bierlich:2014xba,Bierlich:2015rha,Bierlich:2016vgw,Bierlich:2017vhg}. (Appendix~\ref{sec:ropetune}). The predictions are obtained for pp collisions at \thirteen, based on $\inelg$ events.}
    \label{fig:ratioToPion}
\end{figure}

The ratios of the $\pt$-integrated yields of \sigmapm and \xizero hyperons to those of pions are shown in Fig.~\ref{fig:ratioToPion} as a function of the charged-particle pseudorapidity density and they are compared with the ratios measured in pp collisions at $\cms$~=~7~TeV and \pPb collisions at $\snn$ = 5.02 TeV~\cite{Tsallis:1987eu,ALICE:2014zxz,ALICE:2017pgw,ALICE:2020jsh}. They provide insight into the evolution of strangeness production with increasing multiplicity. The results show a smooth increasing trend as a function of multiplicity without energy and collision system dependence. The yields of \sigmapm and \xizero relative to those of pions increase by 60\% and 120\%, respectively, from the lowest to highest multiplicity pp collisions considered in this work.
The increase depends on the strangeness content ($S$) of the resonance; with $\Sigma^{*\pm}$ having $S$=1 and $\Xi^{*0}$ having $S$=2. These results are consistent with previous measurements of ground-state hyperons to pion ratios with ALICE~\cite{ALICE:2019avo}. 
EPOS-LHC and $\pythiaDef$ with Rope shoving predict an increasing trend with multiplicity for both resonances.
EPOS-LHC describes fairly well the measured \sigmapm/$\pi$ ratios, while $\pythiaDef$ Rope overestimates them. Both EPOS-LHC and $\pythiaDef$ Rope tend to overestimate the increasing trend of \xizero/$\pi$ ratios.

\begin{figure}[!htb]
    \centering
    \includegraphics[width=\linewidth]{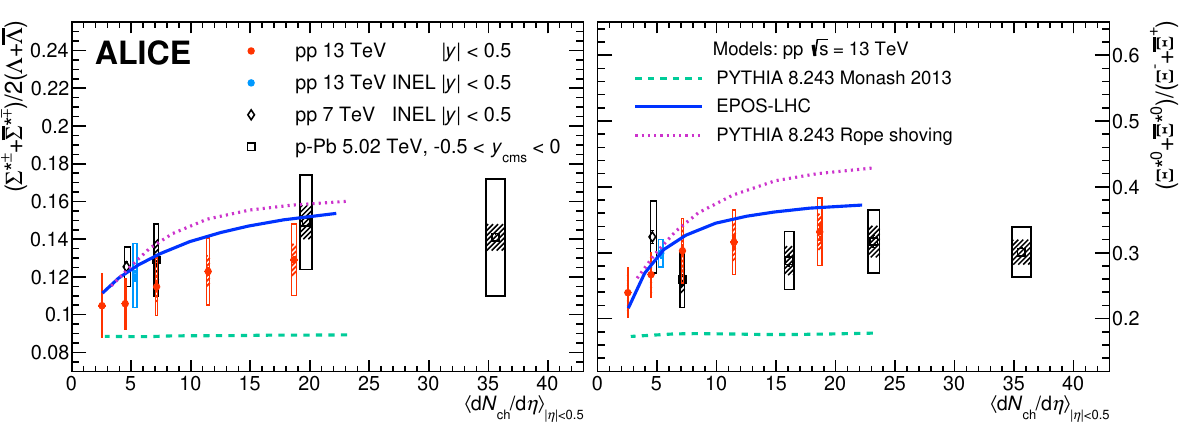}
    \caption{Yield ratio of the resonances to the ground states  having the same quark content as a function of the charged-particle pseudorapidity density for $\sigmapm$ (left) and $\xizero$ (right). The open and shaded boxes represent the total and multiplicity-uncorrelated systematic uncertainties, respectively. The measured points are compared with predictions from different event generators, namely EPOS-LHC~\cite{Pierog:2013ria}, $\pythiaDef$ with Monash 2013 tuning~\cite{Skands:2014pea}, and $\pythiaDef$ with Rope shoving ~\cite{Bierlich:2014xba,Bierlich:2015rha,Bierlich:2016vgw,Bierlich:2017vhg}. (Appendix~\ref{sec:ropetune}). The predictions are obtained for pp collisions at \thirteen, based on $\inelg$ events.}
    \label{fig:ratioToLambdaXi}
\end{figure}

The integrated yield ratios of excited to ground-state hyperons~\cite{ALICE:2014zxz,ALICE:2017pgw,ALICE:2019avo} with the same strangeness content are shown in Fig.~\ref{fig:ratioToLambdaXi}. They are drawn for different collision systems and centre-of-mass energies as a function of \avdndeta. 
The measured \sigmapm/$\Lambda$ and \xizero/$\Xi^{-}$ ratios are compatible either with a flat behaviour as a function of multiplicity or with a mild multiplicity dependence, even though no firm conclusion can be drawn considering the magnitude of the systematic uncertainties and the fact that they are partly uncorrelated across multiplicity intervals.
The EPOS-LHC and $\pythiaDef$ with Rope shoving predict a slight increase of the \sigmapm/$\Lambda$ and \xizero/$\Xi^{-}$ ratios with increasing multiplicity at low multiplicities. 
They describe the data within the experimental uncertainties. 
The $\pythiaDef$ Monash 2013 prediction exhibits a flat behaviour and it underestimates the overall magnitude of the ratios by about a factor of two.
Note that a decreasing trend was observed in the EPOS-LHC model prediction for the K$^{*}/$K ratio ($c\tau$(K$^{*}$) = $4.16$ fm/$c$~\cite{ParticleDataGroup:2022pth}), apparently describing the measurements in \pp and \pPb collisions~\cite{Acharya:2019bli,ALICE:2016sak}.
Despite similar lifetimes, K$^{*}$ and \sigmapm exhibit different yield trends in \pp collisions, hinting at the nuanced interplay of rescattering and regeneration effects within a potential hadronic stage. While the decreasing yield of K$^{*}$ could be attributed to decay product rescattering, the stable yield of \sigmapm might suggest a more pronounced regeneration effect. Conversely, the longer-lived \xizero, seemingly unaffected by these hadronic stage effects, provides a contrasting reference.

\section{Conclusions}

In this article, the $\pt$-differential yields of \sigmapm and \xizero in inelastic \pp collisions at $\cms$ = 13 TeV are reported and compared with previous ALICE measurements in \pp collisions at $\cms$ = 7 TeV, revealing a hardening of the $\pt$ spectra as the collision energy increases. The hardening  is more pronounced for $\pt>2$ \GeVc.
Going from low to high multiplicity events in \pp collisions at $\cms$ = 13 TeV, a clear hardening is observed that affects the shape of the lowest $\pt$ part of the spectra.

The $\pt$-integrated yield, d$N$/d$y$ of $\sigmapm$, and $\xizero$ is found to increase with charged-particle pseudorapidity density, with a trend that does not depend on collision energy and is the same for \pp and \pPb collisions. This is consistent with previous findings at the LHC~\cite{ALICE:2019avo}, highlighting the fact that the mechanisms of particle production are related primarily to the conditions that determine multiplicity. On the other hand, the increasing trend of the \meanpt as a function of \avdndeta in \pp collisions is slightly steeper than the one in \pPb collisions for both $\sigmapm$ and $\xizero$, as observed for charged particles and other light-flavour hadrons in \pp collisions at different energies and in \pPb collisions at $\snn$ = 5.02 TeV.

An increasing trend with multiplicity is found for the $\sigmapm$/$\pi^{\pm}$ and $\xizero$/$\pi^{\pm}$ ratios. The enhancement is more pronounced for $\xizero$ ($S$=2) than $\sigmapm$ ($S$=1), confirming that strangeness enhancement predominantly depends on the strangeness content, rather than on the hyperon mass~\cite{ALICE:2017jyt}. 
The integrated yields of \sigmapm and \xizero show a scaling with multiplicity consistent within uncertainties with that of the ground-state hyperons with the same strangeness content, indicating that the strange-baryon resonance production and its ground state have a similar increase on multiplicity. 
These results, when combined with other resonance studies in small systems, can provide valuable contributions to our understanding of a possible hadronic stage in pp collisions and on the role of rescattering and regeneration effects in such stage.

%%%%%%%%%%%%%%%%%%%%%%%%%%%%%%%%
% end main text 
%%%%%%%%%%%%%%%%%%%%%%%%%%%%%%%%
%%%%% acknowledgements - handled by EB chairs 
\newenvironment{acknowledgement}{\relax}{\relax}
\begin{acknowledgement}
\section*{Acknowledgements}
% add specific acknowledgements here 
% ...but please don't remove the line below: funding agencies
% will be acknowledged with a custom tex file handled by EB chairs after Collab Round 2
% Version: 2023-07-09

The ALICE Collaboration would like to thank all its engineers and technicians for their invaluable contributions to the construction of the experiment and the CERN accelerator teams for the outstanding performance of the LHC complex.
The ALICE Collaboration gratefully acknowledges the resources and support provided by all Grid centres and the Worldwide LHC Computing Grid (WLCG) collaboration.
The ALICE Collaboration acknowledges the following funding agencies for their support in building and running the ALICE detector:
A. I. Alikhanyan National Science Laboratory (Yerevan Physics Institute) Foundation (ANSL), State Committee of Science and World Federation of Scientists (WFS), Armenia;
Austrian Academy of Sciences, Austrian Science Fund (FWF): [M 2467-N36] and Nationalstiftung f\"{u}r Forschung, Technologie und Entwicklung, Austria;
Ministry of Communications and High Technologies, National Nuclear Research Center, Azerbaijan;
Conselho Nacional de Desenvolvimento Cient\'{\i}fico e Tecnol\'{o}gico (CNPq), Financiadora de Estudos e Projetos (Finep), Funda\c{c}\~{a}o de Amparo \`{a} Pesquisa do Estado de S\~{a}o Paulo (FAPESP) and Universidade Federal do Rio Grande do Sul (UFRGS), Brazil;
Bulgarian Ministry of Education and Science, within the National Roadmap for Research Infrastructures 2020-2027 (object CERN), Bulgaria;
Ministry of Education of China (MOEC) , Ministry of Science \& Technology of China (MSTC) and National Natural Science Foundation of China (NSFC), China;
Ministry of Science and Education and Croatian Science Foundation, Croatia;
Centro de Aplicaciones Tecnol\'{o}gicas y Desarrollo Nuclear (CEADEN), Cubaenerg\'{\i}a, Cuba;
Ministry of Education, Youth and Sports of the Czech Republic, Czech Republic;
The Danish Council for Independent Research | Natural Sciences, the VILLUM FONDEN and Danish National Research Foundation (DNRF), Denmark;
Helsinki Institute of Physics (HIP), Finland;
Commissariat \`{a} l'Energie Atomique (CEA) and Institut National de Physique Nucl\'{e}aire et de Physique des Particules (IN2P3) and Centre National de la Recherche Scientifique (CNRS), France;
Bundesministerium f\"{u}r Bildung und Forschung (BMBF) and GSI Helmholtzzentrum f\"{u}r Schwerionenforschung GmbH, Germany;
General Secretariat for Research and Technology, Ministry of Education, Research and Religions, Greece;
National Research, Development and Innovation Office, Hungary;
Department of Atomic Energy Government of India (DAE), Department of Science and Technology, Government of India (DST), University Grants Commission, Government of India (UGC) and Council of Scientific and Industrial Research (CSIR), India;
National Research and Innovation Agency - BRIN, Indonesia;
Istituto Nazionale di Fisica Nucleare (INFN), Italy;
Japanese Ministry of Education, Culture, Sports, Science and Technology (MEXT) and Japan Society for the Promotion of Science (JSPS) KAKENHI, Japan;
Consejo Nacional de Ciencia (CONACYT) y Tecnolog\'{i}a, through Fondo de Cooperaci\'{o}n Internacional en Ciencia y Tecnolog\'{i}a (FONCICYT) and Direcci\'{o}n General de Asuntos del Personal Academico (DGAPA), Mexico;
Nederlandse Organisatie voor Wetenschappelijk Onderzoek (NWO), Netherlands;
The Research Council of Norway, Norway;
Commission on Science and Technology for Sustainable Development in the South (COMSATS), Pakistan;
Pontificia Universidad Cat\'{o}lica del Per\'{u}, Peru;
Ministry of Education and Science, National Science Centre and WUT ID-UB, Poland;
Korea Institute of Science and Technology Information and National Research Foundation of Korea (NRF), Republic of Korea;
Ministry of Education and Scientific Research, Institute of Atomic Physics, Ministry of Research and Innovation and Institute of Atomic Physics and Universitatea Nationala de Stiinta si Tehnologie Politehnica Bucuresti, Romania;
Ministry of Education, Science, Research and Sport of the Slovak Republic, Slovakia;
National Research Foundation of South Africa, South Africa;
Swedish Research Council (VR) and Knut \& Alice Wallenberg Foundation (KAW), Sweden;
European Organization for Nuclear Research, Switzerland;
Suranaree University of Technology (SUT), National Science and Technology Development Agency (NSTDA) and National Science, Research and Innovation Fund (NSRF via PMU-B B05F650021), Thailand;
Turkish Energy, Nuclear and Mineral Research Agency (TENMAK), Turkey;
National Academy of  Sciences of Ukraine, Ukraine;
Science and Technology Facilities Council (STFC), United Kingdom;
National Science Foundation of the United States of America (NSF) and United States Department of Energy, Office of Nuclear Physics (DOE NP), United States of America.
In addition, individual groups or members have received support from:
Czech Science Foundation (grant no. 23-07499S), Czech Republic;
European Research Council, Strong 2020 - Horizon 2020 (grant nos. 950692, 824093), European Union;
ICSC - Centro Nazionale di Ricerca in High Performance Computing, Big Data and Quantum Computing, European Union - NextGenerationEU;
Academy of Finland (Center of Excellence in Quark Matter) (grant nos. 346327, 346328), Finland.

\end{acknowledgement}

%%%%%%%% Bibliography 
\bibliographystyle{utphys}   % Remember we use title in the biblio
\bibliography{bibliography}
%\input {bibliography.tex}  

%%%%%%%%%%%%%%%%%%%%%%%%%%%%%%%%
% Appendices: yours (if any) + authorlist
%%%%%%%%%%%%%%%%%%%%%%%%%%%%%%%%
\newpage
\appendix

\newpage
\section{Fitting functions used in this paper}
\label{sec:fitfunctions}

\paragraph{$\pt$ exponential function}
\begin{equation}
	\frac{\dd^2 N}{\dd\pt\dd y}=A \pt e^{-\pt/T},
\end{equation}
with normalisation factor $A$ and temperature $T$ as fit parameters.

\paragraph{$\mt$ exponential function}
\begin{equation}
	\frac{\dd^2 N}{\dd\pt\dd y}=A \pt e^{-\mt/T},
\end{equation}
where $\mt=\sqrt{\pt^2+m_{0}^2}$ while $m_{0}$ is the rest mass of the particle. The normalisation factor $A$ and temperature $T$ are the fit parameters.

\paragraph{Fermi--Dirac function}\cite{stathandbook}
\begin{equation}
	\frac{\dd^2 N}{\dd\pt\dd y}=\pt A \frac{1}{\rm{e}^{(\sqrt{\it{\pt}^2+\it{m}^2}/\it{T})}+1},
\end{equation}
with $A$ and $T$ as fit parameters and $m$ the mass of the particle under study.

\paragraph{Boltzmann distribution}\cite{stathandbook}
\begin{equation}
  \frac{\dd^2 N}{\dd\pt\dd y}=A\pt\mt e^{-\mt/T},
\end{equation}
where $\mt$ is $\sqrt{\pt^2+m_{0}^2}$ while $m_{0}$ is the rest mass of the particle. The normalisation factor $A$ and temperature $T$ are the fit parameters.

\paragraph{\levi distribution}\cite{Tsallis:1987eu}
\begin{equation}
   \frac{\mathrm{d}^2N}{\mathrm{d}p_\mathrm{T}\mathrm{d}y} = p_\mathrm{T} N' \frac{(n-1)(n-2)}{nC\left[nC+m_0(n-2)\right]} \left[ 1+\frac{\mt-m_{0}}{nC} \right]^{-n} \quad,
\end{equation}

where $\mt=\sqrt{\pt^2+m_{0}^2}$ while $m_{0}$ is the rest mass of the particle, and $n$, $C$ and the integrated yield $N'$ are free parameters of the fit.

\paragraph{Blast-wave distribution}\cite{Schnedermann:1993ws}
\begin{equation}
	\frac{1}{\pt}\frac{\dd^2 N}{\dd\pt\dd y}\propto\int^{R}_{0}rm_{\mathrm{T}}I_0\left(\frac{\pt\sinh\rho}{T_{\mathrm{kin}}}\right)K_1\left(\frac{m_{\mathrm{T}}\cosh\rho}{T_{\mathrm{kin}}}\right)\dd r,
\end{equation}
where $I_0$ and $K_1$ are the modified Bessel functions, $r$ is the distance from the centre of the expanding system, $R$ is the limiting radius of the system expansion, $T_{\mathrm{kin}}$ is the temperature of the kinetic freeze--out and $\rho=\rm{arctanh}\,\beta$ defines the velocity profile.

\newpage
\section{PYTHIA8 Rope Shoving Parameters}
\label{sec:ropetune}

Parameters below are additional parameters to the default Monash tune 2013 (Tune ID = 14 in v8.243).

\begin{verbatim}
MultiPartonInteractions:pT0Ref = 2.15
BeamRemnants:remnantMode = 1
BeamRemnants:saturation = 5
ColourReconnection:mode = 1
ColourReconnection:allowDoubleJunRem = off
ColourReconnection:m0 = 0.3
ColourReconnection:allowJunctions = on
ColourReconnection:junctionCorrection = 1.2
ColourReconnection:timeDilationMode = 2
ColourReconnection:timeDilationPar = 0.18
Ropewalk:RopeHadronization = on
Ropewalk:doShoving = on
Ropewalk:tInit = 1.5
Ropewalk:deltat = 0.05
Ropewalk:tShove 0.1
Ropewalk:gAmplitude = 0.
Ropewalk:doFlavour = on
Ropewalk:r0 = 0.5
Ropewalk:m0 = 0.2
Ropewalk:beta = 0.1
PartonVertex:setVertex = on
PartonVertex:protonRadius = 0.7
PartonVertex:emissionWidth = 0.1
\end{verbatim}

%
%\input{} % put your appendices here (if any)
%

%%%%% Authorlist - please do not touch: handled by EB chairs 
\newpage
\section{The ALICE Collaboration}
\label{app:collab}
% ALICE Collaboration author list for 2023-07-09
\begin{flushleft} 
\small

S.~Acharya\,\orcidlink{0000-0002-9213-5329}\,$^{\rm 128}$, 
D.~Adamov\'{a}\,\orcidlink{0000-0002-0504-7428}\,$^{\rm 87}$, 
G.~Aglieri Rinella\,\orcidlink{0000-0002-9611-3696}\,$^{\rm 33}$, 
M.~Agnello\,\orcidlink{0000-0002-0760-5075}\,$^{\rm 30}$, 
N.~Agrawal\,\orcidlink{0000-0003-0348-9836}\,$^{\rm 52}$, 
Z.~Ahammed\,\orcidlink{0000-0001-5241-7412}\,$^{\rm 136}$, 
S.~Ahmad\,\orcidlink{0000-0003-0497-5705}\,$^{\rm 16}$, 
S.U.~Ahn\,\orcidlink{0000-0001-8847-489X}\,$^{\rm 72}$, 
I.~Ahuja\,\orcidlink{0000-0002-4417-1392}\,$^{\rm 38}$, 
A.~Akindinov\,\orcidlink{0000-0002-7388-3022}\,$^{\rm 142}$
% ...

\section*{Affiliation Notes}

$^{\rm I}$ Deceased\\
$^{\rm II}$ Also at: Max-Planck-Institut fur Physik, Munich, Germany\\
$^{\rm III}$ Also at: Italian National Agency for New Technologies, Energy and Sustainable Economic Development (ENEA), Bologna, Italy\\
$^{\rm IV}$ Also at: Dipartimento DET del Politecnico di Torino, Turin, Italy\\
$^{\rm V}$ Also at: Yildiz Technical University, Istanbul, T\"{u}rkiye\\
$^{\rm VI}$ Also at: Department of Applied Physics, Aligarh Muslim University, Aligarh, India\\
$^{\rm VII}$ Also at: Institute of Theoretical Physics, University of Wroclaw, Poland\\
$^{\rm VIII}$ Also at: An institution covered by a cooperation agreement with CERN\\

\section*{Collaboration Institutes}

$^{1}$ A.I. Alikhanyan National Science Laboratory (Yerevan Physics Institute) Foundation, Yerevan, Armenia\\
$^{2}$ AGH University of Krakow, Cracow, Poland\\
$^{3}$ Bogolyubov Institute for Theoretical Physics, National Academy of Sciences of Ukraine, Kiev, Ukraine\\
$^{4}$ Bose Institute, Department of Physics  and Centre for Astroparticle Physics and Space Science (CAPSS), Kolkata, India\\
$^{5}$ California Polytechnic State University, San Luis Obispo, California, United States\\
$^{6}$ Central China Normal University, Wuhan, China\\
$^{7}$ Centro de Aplicaciones Tecnol\'{o}gicas y Desarrollo Nuclear (CEADEN), Havana, Cuba\\
$^{8}$ Centro de Investigaci\'{o}n y de Estudios Avanzados (CINVESTAV), Mexico City and M\'{e}rida, Mexico\\
$^{9}$ Chicago State University, Chicago, Illinois, United States\\
$^{10}$ China Institute of Atomic Energy, Beijing, China\\
$^{11}$ China University of Geosciences, Wuhan, China\\
$^{12}$ Chungbuk National University, Cheongju, Republic of Korea\\
$^{13}$ Comenius University Bratislava, Faculty of Mathematics, Physics and Informatics, Bratislava, Slovak Republic\\
$^{14}$ COMSATS University Islamabad, Islamabad, Pakistan\\
$^{15}$ Creighton University, Omaha, Nebraska, United States\\
$^{16}$ Department of Physics, Aligarh Muslim University, Aligarh, India\\
$^{17}$ Department of Physics, Pusan National University, Pusan, Republic of Korea\\
$^{18}$ Department of Physics, Sejong University, Seoul, Republic of Korea\\
$^{19}$ Department of Physics, University of California, Berkeley, California, United States\\
$^{20}$ Department of Physics, University of Oslo, Oslo, Norway\\
$^{21}$ Department of Physics and Technology, University of Bergen, Bergen, Norway\\
$^{22}$ Dipartimento di Fisica, Universit\`{a} di Pavia, Pavia, Italy\\
$^{23}$ Dipartimento di Fisica dell'Universit\`{a} and Sezione INFN, Cagliari, Italy\\
$^{24}$ Dipartimento di Fisica dell'Universit\`{a} and Sezione INFN, Trieste, Italy\\
$^{25}$ Dipartimento di Fisica dell'Universit\`{a} and Sezione INFN, Turin, Italy\\
$^{26}$ Dipartimento di Fisica e Astronomia dell'Universit\`{a} and Sezione INFN, Bologna, Italy\\
$^{27}$ Dipartimento di Fisica e Astronomia dell'Universit\`{a} and Sezione INFN, Catania, Italy\\
$^{28}$ Dipartimento di Fisica e Astronomia dell'Universit\`{a} and Sezione INFN, Padova, Italy\\
$^{29}$ Dipartimento di Fisica `E.R.~Caianiello' dell'Universit\`{a} and Gruppo Collegato INFN, Salerno, Italy\\
$^{30}$ Dipartimento DISAT del Politecnico and Sezione INFN, Turin, Italy\\
$^{31}$ Dipartimento di Scienze MIFT, Universit\`{a} di Messina, Messina, Italy\\
$^{32}$ Dipartimento Interateneo di Fisica `M.~Merlin' and Sezione INFN, Bari, Italy\\
$^{33}$ European Organization for Nuclear Research (CERN), Geneva, Switzerland\\
$^{34}$ Faculty of Electrical Engineering, Mechanical Engineering and Naval Architecture, University of Split, Split, Croatia\\
$^{35}$ Faculty of Engineering and Science, Western Norway University of Applied Sciences, Bergen, Norway\\
$^{36}$ Faculty of Nuclear Sciences and Physical Engineering, Czech Technical University in Prague, Prague, Czech Republic\\
$^{37}$ Faculty of Physics, Sofia University, Sofia, Bulgaria\\
$^{38}$ Faculty of Science, P.J.~\v{S}af\'{a}rik University, Ko\v{s}ice, Slovak Republic\\
$^{39}$ Frankfurt Institute for Advanced Studies, Johann Wolfgang Goethe-Universit\"{a}t Frankfurt, Frankfurt, Germany\\
$^{40}$ Fudan University, Shanghai, China\\
$^{41}$ Gangneung-Wonju National University, Gangneung, Republic of Korea\\
$^{42}$ Gauhati University, Department of Physics, Guwahati, India\\
$^{43}$ Helmholtz-Institut f\"{u}r Strahlen- und Kernphysik, Rheinische Friedrich-Wilhelms-Universit\"{a}t Bonn, Bonn, Germany\\
$^{44}$ Helsinki Institute of Physics (HIP), Helsinki, Finland\\
$^{45}$ High Energy Physics Group,  Universidad Aut\'{o}noma de Puebla, Puebla, Mexico\\
$^{46}$ Horia Hulubei National Institute of Physics and Nuclear Engineering, Bucharest, Romania\\
$^{47}$ HUN-REN Wigner Research Centre for Physics, Budapest, Hungary\\
$^{48}$ Indian Institute of Technology Bombay (IIT), Mumbai, India\\
$^{49}$ Indian Institute of Technology Indore, Indore, India\\
$^{50}$ INFN, Laboratori Nazionali di Frascati, Frascati, Italy\\
$^{51}$ INFN, Sezione di Bari, Bari, Italy\\
$^{52}$ INFN, Sezione di Bologna, Bologna, Italy\\
$^{53}$ INFN, Sezione di Cagliari, Cagliari, Italy\\
$^{54}$ INFN, Sezione di Catania, Catania, Italy\\
$^{55}$ INFN, Sezione di Padova, Padova, Italy\\
$^{56}$ INFN, Sezione di Pavia, Pavia, Italy\\
$^{57}$ INFN, Sezione di Torino, Turin, Italy\\
$^{58}$ INFN, Sezione di Trieste, Trieste, Italy\\
$^{59}$ Inha University, Incheon, Republic of Korea\\
$^{60}$ Institute for Gravitational and Subatomic Physics (GRASP), Utrecht University/Nikhef, Utrecht, Netherlands\\
$^{61}$ Institute of Experimental Physics, Slovak Academy of Sciences, Ko\v{s}ice, Slovak Republic\\
$^{62}$ Institute of Physics, Homi Bhabha National Institute, Bhubaneswar, India\\
$^{63}$ Institute of Physics of the Czech Academy of Sciences, Prague, Czech Republic\\
$^{64}$ Institute of Space Science (ISS), Bucharest, Romania\\
$^{65}$ Institut f\"{u}r Kernphysik, Johann Wolfgang Goethe-Universit\"{a}t Frankfurt, Frankfurt, Germany\\
$^{66}$ Instituto de Ciencias Nucleares, Universidad Nacional Aut\'{o}noma de M\'{e}xico, Mexico City, Mexico\\
$^{67}$ Instituto de F\'{i}sica, Universidade Federal do Rio Grande do Sul (UFRGS), Porto Alegre, Brazil\\
$^{68}$ Instituto de F\'{\i}sica, Universidad Nacional Aut\'{o}noma de M\'{e}xico, Mexico City, Mexico\\
$^{69}$ iThemba LABS, National Research Foundation, Somerset West, South Africa\\
$^{70}$ Jeonbuk National University, Jeonju, Republic of Korea\\
$^{71}$ Johann-Wolfgang-Goethe Universit\"{a}t Frankfurt Institut f\"{u}r Informatik, Fachbereich Informatik und Mathematik, Frankfurt, Germany\\
$^{72}$ Korea Institute of Science and Technology Information, Daejeon, Republic of Korea\\
$^{73}$ KTO Karatay University, Konya, Turkey\\
$^{74}$ Laboratoire de Physique Subatomique et de Cosmologie, Universit\'{e} Grenoble-Alpes, CNRS-IN2P3, Grenoble, France\\
$^{75}$ Lawrence Berkeley National Laboratory, Berkeley, California, United States\\
$^{76}$ Lund University Department of Physics, Division of Particle Physics, Lund, Sweden\\
$^{77}$ Nagasaki Institute of Applied Science, Nagasaki, Japan\\
$^{78}$ Nara Women{'}s University (NWU), Nara, Japan\\
$^{79}$ National and Kapodistrian University of Athens, School of Science, Department of Physics , Athens, Greece\\
$^{80}$ National Centre for Nuclear Research, Warsaw, Poland\\
$^{81}$ National Institute of Science Education and Research, Homi Bhabha National Institute, Jatni, India\\
$^{82}$ National Nuclear Research Center, Baku, Azerbaijan\\
$^{83}$ National Research and Innovation Agency - BRIN, Jakarta, Indonesia\\
$^{84}$ Niels Bohr Institute, University of Copenhagen, Copenhagen, Denmark\\
$^{85}$ Nikhef, National institute for subatomic physics, Amsterdam, Netherlands\\
$^{86}$ Nuclear Physics Group, STFC Daresbury Laboratory, Daresbury, United Kingdom\\
$^{87}$ Nuclear Physics Institute of the Czech Academy of Sciences, Husinec-\v{R}e\v{z}, Czech Republic\\
$^{88}$ Oak Ridge National Laboratory, Oak Ridge, Tennessee, United States\\
$^{89}$ Ohio State University, Columbus, Ohio, United States\\
$^{90}$ Physics department, Faculty of science, University of Zagreb, Zagreb, Croatia\\
$^{91}$ Physics Department, Panjab University, Chandigarh, India\\
$^{92}$ Physics Department, University of Jammu, Jammu, India\\
$^{93}$ Physics Program and International Institute for Sustainability with Knotted Chiral Meta Matter (SKCM2), Hiroshima University, Hiroshima, Japan\\
$^{94}$ Physikalisches Institut, Eberhard-Karls-Universit\"{a}t T\"{u}bingen, T\"{u}bingen, Germany\\
$^{95}$ Physikalisches Institut, Ruprecht-Karls-Universit\"{a}t Heidelberg, Heidelberg, Germany\\
$^{96}$ Physik Department, Technische Universit\"{a}t M\"{u}nchen, Munich, Germany\\
$^{97}$ Politecnico di Bari and Sezione INFN, Bari, Italy\\
$^{98}$ Research Division and ExtreMe Matter Institute EMMI, GSI Helmholtzzentrum f\"ur Schwerionenforschung GmbH, Darmstadt, Germany\\
$^{99}$ Saga University, Saga, Japan\\
$^{100}$ Saha Institute of Nuclear Physics, Homi Bhabha National Institute, Kolkata, India\\
$^{101}$ School of Physics and Astronomy, University of Birmingham, Birmingham, United Kingdom\\
$^{102}$ Secci\'{o}n F\'{\i}sica, Departamento de Ciencias, Pontificia Universidad Cat\'{o}lica del Per\'{u}, Lima, Peru\\
$^{103}$ Stefan Meyer Institut f\"{u}r Subatomare Physik (SMI), Vienna, Austria\\
$^{104}$ SUBATECH, IMT Atlantique, Nantes Universit\'{e}, CNRS-IN2P3, Nantes, France\\
$^{105}$ Sungkyunkwan University, Suwon City, Republic of Korea\\
$^{106}$ Suranaree University of Technology, Nakhon Ratchasima, Thailand\\
$^{107}$ Technical University of Ko\v{s}ice, Ko\v{s}ice, Slovak Republic\\
$^{108}$ The Henryk Niewodniczanski Institute of Nuclear Physics, Polish Academy of Sciences, Cracow, Poland\\
$^{109}$ The University of Texas at Austin, Austin, Texas, United States\\
$^{110}$ Universidad Aut\'{o}noma de Sinaloa, Culiac\'{a}n, Mexico\\
$^{111}$ Universidade de S\~{a}o Paulo (USP), S\~{a}o Paulo, Brazil\\
$^{112}$ Universidade Estadual de Campinas (UNICAMP), Campinas, Brazil\\
$^{113}$ Universidade Federal do ABC, Santo Andre, Brazil\\
$^{114}$ Universitatea Nationala de Stiinta si Tehnologie Politehnica Bucuresti, Bucharest, Romania\\
$^{115}$ University of Cape Town, Cape Town, South Africa\\
$^{116}$ University of Derby, Derby, United Kingdom\\
$^{117}$ University of Houston, Houston, Texas, United States\\
$^{118}$ University of Jyv\"{a}skyl\"{a}, Jyv\"{a}skyl\"{a}, Finland\\
$^{119}$ University of Kansas, Lawrence, Kansas, United States\\
$^{120}$ University of Liverpool, Liverpool, United Kingdom\\
$^{121}$ University of Science and Technology of China, Hefei, China\\
$^{122}$ University of South-Eastern Norway, Kongsberg, Norway\\
$^{123}$ University of Tennessee, Knoxville, Tennessee, United States\\
$^{124}$ University of the Witwatersrand, Johannesburg, South Africa\\
$^{125}$ University of Tokyo, Tokyo, Japan\\
$^{126}$ University of Tsukuba, Tsukuba, Japan\\
$^{127}$ Universit\"{a}t M\"{u}nster, Institut f\"{u}r Kernphysik, M\"{u}nster, Germany\\
$^{128}$ Universit\'{e} Clermont Auvergne, CNRS/IN2P3, LPC, Clermont-Ferrand, France\\
$^{129}$ Universit\'{e} de Lyon, CNRS/IN2P3, Institut de Physique des 2 Infinis de Lyon, Lyon, France\\
$^{130}$ Universit\'{e} de Strasbourg, CNRS, IPHC UMR 7178, F-67000 Strasbourg, France, Strasbourg, France\\
$^{131}$ Universit\'{e} Paris-Saclay, Centre d'Etudes de Saclay (CEA), IRFU, D\'{e}partment de Physique Nucl\'{e}aire (DPhN), Saclay, France\\
$^{132}$ Universit\'{e}  Paris-Saclay, CNRS/IN2P3, IJCLab, Orsay, France\\
$^{133}$ Universit\`{a} degli Studi di Foggia, Foggia, Italy\\
$^{134}$ Universit\`{a} del Piemonte Orientale, Vercelli, Italy\\
$^{135}$ Universit\`{a} di Brescia, Brescia, Italy\\
$^{136}$ Variable Energy Cyclotron Centre, Homi Bhabha National Institute, Kolkata, India\\
$^{137}$ Warsaw University of Technology, Warsaw, Poland\\
$^{138}$ Wayne State University, Detroit, Michigan, United States\\
$^{139}$ Yale University, New Haven, Connecticut, United States\\
$^{140}$ Yonsei University, Seoul, Republic of Korea\\
$^{141}$  Zentrum  f\"{u}r Technologie und Transfer (ZTT), Worms, Germany\\
$^{142}$ Affiliated with an institute covered by a cooperation agreement with CERN\\
$^{143}$ Affiliated with an international laboratory covered by a cooperation agreement with CERN.\\

\end{flushleft} 

\end{document}